<u>*Ideal Mixing of Paraelectric and Ferroelectric Nematic Phases*</u>
<u>*in Liquid Crystals of Distinct Molecular Species*</u>

Xi Chen,[1] Zhecong Zhu,[1] Mitchell J. Magrini,[2] Eva Korblova,[2] Cheol S. Park,[1]
Matthew A. Glaser,[1] Joseph E. Maclennan,[1] David M. Walba,[2] Noel A. Clark[1]*

*[1]Department of Physics and Soft Materials Research Center,*
*University of Colorado, Boulder, CO 80309, USA*

*[2]Department of Chemistry and Soft Materials Research Center,*
*University of Colorado, Boulder, CO 80309, USA*

*Abstract*

The organic mesogens RM734 and DIO are members of separate molecular families featuring distinct molecular structures. These families, at the present time, are the only ones known to exhibit a ferroelectric nematic liquid crystal (LC) phase. Here we present an experimental investigation of the phase diagram and electro-optics of binary mixtures of RM734 and DIO. We observe paraelectric nematic (N) and ferroelectric nematic ($N_F$) phases in both materials, each of which exhibits complete miscibility across the phase diagram, showing that the paraelectric and ferroelectric are the same phases in RM734 as in DIO. Remarkably, these molecules form ideal mixtures with respect to both the paraelectric-ferroelectric nematic phase behavior and the ferroelectric polarization density of the mixtures, the principal order parameter of the transition. Ideal mixing is also manifested in the orientational viscosity, and the onset of glassy dynamics at low temperature. This behavior is attributable in part to the similarity of their overall molecular shape and net longitudinal dipole moment (~ 11 Debye), and to a common tendency for head-to-tail molecular association. In contrast, the significant difference in molecular structures leads to poor solubility in the crystal phases, enhancing the stability of the ferroelectric nematic phase at low temperature in the mixtures and making possible room temperature electro-optic effects. In the mixtures with excess DIO, an intermediate phase appears via an ultraweak, first-order transition from the N phase, in a narrow temperature range between the paraelectric and ferroelectric nematics.

## INTRODUCTION

Ferroelectricity in liquids was predicted in the 1910s by P. Debye [1] and M. Born [2], who applied the Langevin-Weiss model of ferromagnetism to the orientational ordering of molecular electric dipoles. A century later, in 2017, two groups independently reported novel nematic phases of polar molecules, the antiferroelectric splay nematic in the molecule RM734 [3,4,5] and a "ferroelectric-like" phase in the molecule DIO [6]. Ferroelectricity has subsequently been demonstrated in RM734 [7] and in DIO, as confirmed in our own experiments and by Li et al. [8]. The serendipity of this development is remarkable since, as is evident from *Fig. 1*, RM734 and DIO are members of separate molecular families with distinctly different molecular structures. Novel polar nematics have been independently observed in close homologs and mixtures within these families [3,8,9,10,11,12]. On the other hand, DIO and RM734 have similar molecular shape and size, and both molecules have longitudinal molecular dipole moments of ~ 11 Debye, similarities which could favor their miscibility in an $N_F$ phase.

These observations motivated us to pursue the study of the interactions between these distinct molecular species in the context of nematic ferroelectricity. Here we present an experimental investigation of the phase and electro-optic behavior of binary DIO/RM734 mixtures. Similarities in the optical textures, calorimetry, and second-harmonic generation of RM734 and DIO were recently reported by Li et al. [8], but the question of whether the ferroelectric nematics identified in RM734 and DIO are the same phase can be answered unambiguously only by investigating binary miscibility [13]. The phenomenology of mixing these materials in their ferroelectric nematic phases was unknown, but, because of their chemical differences, it appeared unlikely that these molecules would be very miscible in their crystal phases, promising an opportunity to suppress crystallization and achieve room-temperature, ferroelectric mixtures. In relation to this study, we also investigated the M2 phase originally reported in DIO [6] but not structurally characterized. Synchrotron-based microbeam small-angle x-ray scattering (SAXS) and electro-optic polarized light microscopy show the M2 to be a lamellar, density-modulated antiferroelectric LC having a nematic director parallel to the layer planes, a phase which we term Smectic Z (SmZ$_A$) in work reported elsewhere [14].

Nematic ferroelectricity presents opportunities for novel liquid crystal science and technology thanks to its unique combination of macroscopic polar ordering and fluidity. The ferroelectric nematic ($N_F$) phase of RM734 shows a rapid electro-optic response at high temperature in the $N_F$ range [15] but exhibits crystallization and a viscosity that grows strongly on slow cooling. The room temperature $N_F$ phase that is obtained by quenching, on the other hand, is glassy. In the applications development of liquid crystal technologies, the exploration of mixtures is a time-honored and highly successful approach to addressing issues such as eliminating crystallization [16], expanding phase ranges [17], and tuning liquid crystal properties [18]. Studies of mixtures are also key to advancing liquid crystal science, providing a way to test structural models of phases by continuously varying the composition [19], to find phases in mixtures that are not exhibited by any of their components [20,21], and to discover new phases [22].

## *RESULTS*

*Phase Diagram* – RM734 and DIO were synthesized using respectively the schemes published in [7] and shown in *Fig. S1*. The phase diagram for different weight percent (wt%) of DIO in the mixture, *c*, determined upon slow cooling from the isotropic (Iso) phase using polarized light microscopy (PLM) in transmission, differential scanning calorimetry (DSC), polarization measurement, and SAXS experiments, is shown in *Fig. 2*. A well-defined phase front passing through the cell was observed in the microscope at each of the transitions, allowing an accurate determination of the transition temperatures. The observed transition temperatures of the neat components agree well with the published values [3,6]. Upon cooling the mixtures, we observed three different liquid crystal phases: paraelectric nematic (N); antiferroelectric smectic Z (SmZ$_A$); and ferroelectric nematic (N$_F$). We refer to the N phase as "paraelectric" using the standard condensed matter terminology for the disordered phase above a ferroelectric or antiferroelectric. At the lowest temperatures, we observed crystal phases, but their properties and composition were not investigated. The Iso, N, and N$_F$ phases appeared in continuous fashion across the entire phase diagram, indicating complete miscibility of the two components at all concentrations in these phases. According to the miscibility rule [13], this observation answers the question posed above, indicating that the Iso, N, and N$_F$ phases in RM734 are the same phases as in DIO. The phase range of the SmZ$_A$ (denoted M2 in [6]), about 15 ℃ in neat DIO, is reduced with decreasing DIO concentration, disappearing below *c* ~ 50 wt%. As is evident from *Fig. 2*, the phase boundary into the N$_F$ is sloped and linear in *c*, suggesting ideal mixing behavior at this transition, as discussed in detail below.

Phase properties of note can be summarized as follows.

•*Isotropic Phase (Iso)* – The Iso phase shows extinction of transmitted light between crossed polarizer and analyzer and exhibits no detectable response to applied fields of up to 100 V/mm.

•*Paraelectric Nematic Phase (N)* – As expected, the N phase adopts planar alignment with a uniform, in-plane director field, *n(r)*, parallel to the buffing direction. Excellent extinction is achieved when the cell is oriented with the director parallel to the polarizer or analyzer (*Fig. 4A,S3B*). Freedericksz transitions, driven by a 200 Hz square-wave field, were observed in the N phase (*Fig. S3C*), with in-plane fields generating twist deformation of the director [RMS threshold voltage $V_{th}^{T} = 2\pi(D/d)\sqrt{K_T/(\varepsilon_0\Delta\varepsilon)}$], and fields normal to the cell plates generating splay-bend deformation [RMS threshold voltage $V_{th}^{S} = 2\pi\sqrt{K_S/(\varepsilon_0\Delta\varepsilon)}$] [23], where *D* is the in-plane electrode gap, $K_T$ and $K_S$ are the twist and splay Frank elastic constants, and $\Delta\varepsilon$ is the low-frequency dielectric anisotropy. The temperature dependence of the threshold voltages in RM734 and DIO is shown in ***Fig. S4***. RM734 and DIO exhibit typical Freedericksz-like, thresholded orientational response to applied field. Both RM734 and DIO exhibit a general increase in $\Delta\varepsilon$ on cooling. In RM734, however, the Freedericksz thresholds are lower and decrease on approaching the transition to the $N_F$ phase due to the strong pretransitional growth of $\Delta\varepsilon$ and decrease of $K_S$ [4,5], phenomena which do not occur in the N phase of DIO because the N approaches an antiferroelectric phase. The observed splay threshold voltage of RM734 agrees well with $V_{th}^S$ calculated using the $K_S$ and $\Delta\varepsilon$ values given in Refs. [4,5]. Comparison of the measurements with the theoretical twist/splay threshold ratio $V_{th}^T / V_{th}^S = (D / d)\sqrt{K_T / K_S} = (1\,\text{mm} / 3.5\,\mu\text{m})\sqrt{K_T / K_S} = 286\sqrt{K_T / K_S}$ enables an estimate of the elastic constant ratio $K_T/K_S$. The dashed lines in ***Fig. S4*** give $V_{th}^S$ scaled up by 286, showing that $K_S > K_T$ in both materials, except near the N–$N_F$ transition in RM734, due to the pretransitional decrease of $K_S$ [4,5]. In DIO, $K_S \sim 10\,K_T$ over most of the nematic range.

• _Lamellar Antiferroelectric LC with In-Plane Nematic Director (SmZ$_A$)_ – We have recently determined, using non-resonant SAXS on magnetically aligned capillaries and PLM electro-optics of aligned cells, that the previously reported [6] but structurally uncharacterized M2 phase of DIO is a layered (density-modulated), antiferroelectric LC, which we name the smectic Z$_A$ (SmZ$_A$), comprising a periodic array of 9-nm thick, polar layers with alternating polarization and a nematic director parallel to the layer planes [14]. In cells, these layers fill three-dimensional space in well-defined, smectic-like geometries, with the layers either parallel or normal to the plates. Upon cooling from the N phase in a cell with buffed polyimide surfaces, the layers grow in normal to the plates, suppressing the in-plane twist Freedericksz response to fields applied in the plane of the cell, but preserving the splay-bend Freedericksz transition for fields applied normal to the plates. These features make it easy to distinguish the SmZ$_A$ from both the N and $N_F$ phases.

• _Ferroelectric Nematic Phase (N$_F$)_ – The textural evolution of the $N_F$ phase obtained on cooling a uniform domain of the SmZ$_A$ phase in neat DIO is shown in the PLM images of ***Figs. 3A-D***, and that on cooling a uniform domain of the N phase in the $c = 40$ wt% DIO mixture in ***Fig. 4***, both in the absence of field. In addition to undergoing characteristic optical changes, the transition to the $N_F$ is marked by an increase in the threshold for the splay-bend Freedericksz transition, from a few tenths of a volt in the N phase to more than one hundred volts in the $N_F$ phase, a result of the large electrostatic energy cost of rotating the ferroelectric polarization ***P*** in an initially planar cell to give a component normal to the cell plates [7]. At the same time, the threshold field for in-plane field-induced twist is reduced by about a factor of 1000 because of the development of ferroelectric coupling between ***P*** and ***E*** [7], giving extreme electro-optic responsivity in very weak applied in-plane electric fields (in the 0.1 to 1 V/mm range) in all of the $N_F$ director states shown. At the N to $N_F$ transition, a front characterized by local orientation fluctuations passes through the cell (***Fig. 4B***), leaving behind a smooth, planar texture in the $N_F$ phase, retaining the uniform director field of the N (and SmZ$_A$) phases with $n$ parallel to the buffing axis (the $U$ state) for several degrees into the N$_F$ phase. On further cooling, however, a structural transition occurs, with distinct transition lines ($\pi$-twist disclination lines formed at one surface) passing laterally across the cell and mediating the formation of left- and right-handed ($LH$ and $RH$) $\pi$-twist domains, which are non-extinguishing and are themselves separated by another kind of distinct line defect ($2\pi$-twist lines in the cell midplane), seen in *Figs. 3A,4C*. These twisted states appear optically identical between crossed polarizer and analyzer, but their equivalence is lost when the polarizers are decrossed, with the LH and RH domains exhibiting distinct colors that are exchanged when the polarizers are decrossed the other way (*Figs. 3B-D*). This behavior, first observed in *ANTIPOLAR* cells of neat RM734 [15], indicates a spontaneous transformation of the uniform nematic director/polarization state into a $\pi$-twisted state by passage of a $\pi$ twist line. This observation is critical and unambiguous confirmative evidence for nematic ferroelectricity: a spontaneous uniform to $\pi$-twisted transition in an antiparallel-buffed cell in the N$_F$ phase is a uniquely ferroelectric nematic phenomenon, requiring not only macroscopic polar ordering of the bulk LC but also polar coupling to a macroscopically polar surface. The *LH* and *RH* $\pi$-twist states support, respectively, a half-turn of a left- or right-handed director helix and are separated by topological $2\pi$-twist lines. The two twist states have opposite net polarization, normal to the buffing axis, so that reversing an electric field applied in this direction can be used to switch between them, as shown in *Figs. 3E-G*. Such $\pi$-twist states are observed in all of the mixtures and in neat DIO, and show behavior qualitatively similar to that observed in neat RM734 [7,15], leading us to conclude that the N$_F$ is continuous across the phase diagram. Li et al. have also noted similarities in RM734 and DIO textures observed in random-planar cells [8].

*Transition to the N$_F$ Phase* – This transition is distinctly different at the two ends of the phase diagram. In RM734-rich mixtures, upon approaching the N – N$_F$ transition, we observe the random pattern of fluctuating polar domains, as previously reported in neat RM734 [4,7] and shown in *Fig. 4B*. These domains are extended along the director orientation, $n$, in a manifestation of the electrostatic suppression of longitudinal fluctuations of the polarization $P$ [7]. The details of coarsening of these domains upon cooling into the N$_F$ depends on the surface conditions, but in few-micron thick, rubbed polyimide cells the domains typically grow to several microns in size, eliminating defects in the texture and forming large, uniform monodomain and then twisted states [15]. In thicker or more weakly aligned cells, the length scale of the coarsening domains has been observed to increase continuously on cooling to millimeter dimensions [24], with irregular, macroscopic patterns of reversed polarization extended along the director, as seen in RM734 [7,8,11,24] and in a homolog of DIO that transitions directly from N to N$_F$ [8]. In typical test cells, the uniform N$_F$ domains are separated either by pure polarization reversal walls or by splay-bend walls [7,8]. This behavior, along with the ferroelectric uniform and twisted states observed in DIO, suggest that the $N_F$ phase observed in the mixtures is the same as that in RM734 and DIO.

We note that we have not observed periodic birefringent (splay nematic) stripes with ~9 μm spacing of the kind previously reported in the $N_F$ phase of the RM734 family by Mandle and coworkers [5,11], not in the $N_F$ phase of any preparations of RM734, DIO or their mixtures, in any standard thin cells, capillaries, or thicker (10–50 μm) cells with planar or random-planar alignment.

At the DIO end of the phase diagram, the transition sequence on cooling is first $N - SmZ_A$ and then $SmZ_A - N_F$.  These transitions are weakly first-order, and the phases grow in as optically distinct, uniform domains upon cooling, without the dramatic polar fluctuations seen at the N – $N_F$ transition.  We attribute this to the antiferroelectric ordering of the $SmZ_A$ phase.  Over most of the $SmZ_A$ phase, in-plane reorientation of the director field is strongly suppressed but at lower temperatures, approaching the transition to the $N_F$ phase, ferroelectric fluctuations appear and the susceptibility for field-induced reorientation increases, to be discussed in a later publication.

_Ferroelectric Polarization_ – A typical set of polarization current, $i(t)$, vs. time measurements in response to a 50 Hz, 104 V/mm square-wave, in-plane applied field at different temperatures, in this case for the $c$ = 90 wt% DIO mixture, is plotted in **Fig. 5A**, with additional data included in **Fig. S5**.  This weak applied field is large enough to reverse the polarization in the $N_F$.  In the Iso, N and $SmZ_A$ phases, the current consists only of a signal that peaks shortly after sign reversal of the applied voltage and then decays exponentially.  This signal corresponds to the RC-circuit linear response of the cell and series resistance, giving the initial upward curvature of the measured $P$, due to increasing $\varepsilon$ in the N phase as the $N_F$ is approached in $T$.  At the transition to the $N_F$, a much larger current signal, resulting from the reversal of spontaneous polarization in the sample, appears at longer times.  This current peak is integrated in time to obtain the net charge flow $Q = \int i(t)dt$ and the corresponding charge density $Q/2A$, shown in **Figs. 5B,S5**, where $A$ is the cross-section of the liquid crystal sample in the plane normal to the applied field midway between the two electrodes.

Initial qualitative observations showed that the width in time of the current peak increased dramatically with decreasing $T$ (**Fig. S5**), motivating the choice of square wave driving in order to minimize the temporal width of the current response.  Thus, at high temperatures, polarization reversal is completed during the available 10 msec integration time between applied field reversals.  In this regime, denoted by the filled square symbols in **Fig 5B**, the quantity $Q/2A$ is equivalent to the bulk $N_F$ polarization, with $P(c,T) = Q(c,T)/2A$.  We find that in this temperature range, the dependence of $P(T)$ on temperature is quite similar in all of the mixtures: on approaching the transition to the $N_F$ phase, the polarization increases sigmoidally, with an initial upward curvature reflecting the pre-transitional increase in dielectric constant in the N (or SmZ$_A$) phase observed in both RM734 [4,5] and DIO [6, 8].

However, as **Figs. 5A,C** show, the full width at half-maximum of the current peak associated with polarization reversal, $\tau_R$, increases rapidly on cooling. Below a temperature $T_{sat}$, even with square wave driving, polarization reversal cannot be completed within the available 10 msec time window and the measured $Q(c,T)/2A$ values decrease rapidly with decreasing $T$ from their maximum value of $P_{sat}$ (solid circles in **Fig. 5B**). Test experiments with longer integration times confirm that these data do not reflect the true polarization, which increases by a few percent above $P_{sat}(c)$ as $T$ is lowered, as observed in RM734 and confirmed in atomistic simulations of RM734 [7]. In the following analysis of the time reversal dynamics, we approximate the polarization at temperatures below $T_{sat}$ simply by $P_{sat}(c)$. Under this assumption, the time reversal dynamics can be used to obtain measurement of the orientational viscosity at temperatures $T < T_{sat}$.

The saturation value of the polarization is similar in all of the mixtures, $P_{sat} \sim 6$ µC/cm², as seen in **Fig. 5B**, with $P_{sat}(c)$ decreasing slightly from the RM734-rich to the DIO-rich end, as shown in **Fig. 5D**.

At low temperatures, the polarization measurements were made approximately 30 minutes apart to allow time for possible crystallization. The inset in **Fig. 5A** shows the current response following field reversal of the $c = 90\%$ DIO mixture at the lowest temperatures. In this mixture, crystallization occurred between the $T = 26.5$ °C and $T = 25.5$ °C scans, causing a precipitous drop in the integrated current, as is evident from the plot. In general, crystallization is observed on cooling in mixtures at both ends of the phase diagram, near $c \sim 0$ and $c \sim 100\%$, as crystallization becomes thermodynamically favorable at higher temperatures where the fluid N$_F$ phase still has relatively low viscosity (gray Xtal regions in **Fig. 2**). Interestingly, crystallization is largely suppressed in the $c = 90\%$ DIO mixture while the square-wave polarization reversal field is applied, but the sample crystallizes within about 1 hr in the absence of field at $T = 25$ °C. The polarization reversal time was measured as a function of temperature for all the mixtures, with the results shown in **Fig. 5C**.

For $T < T_{sat}$, the measured $Q(T)$ data, although not giving the full polarization, can be used to provide an upper bound estimate $f(T) = (Q(T)/2A)/P_{sat}(c)$ of the fraction of the polarization that has reoriented within the 10 msec integration window, under the assumption that for $T < T_{sat}$, $P(c,T) = P_{sat}(c)$. The precipitous reduction in the measured charge density at the lowest temperatures coincides with a sudden increase in switching time, which results from a rapid rise in the effective orientational viscosity on approaching the glassy state. In the middle range of concentrations, there is no evidence of crystallization, which is impeded at low $T$ by the high viscosity and by freezing-point depression, with cooling resulting instead in a glassy state. Crystallization could be suppressed at all DIO concentrations by rapid cooling, enabling any mixture to be quenched into a room-temperature glass.

*Polarization Reorientation Dynamics: Orientational Viscosity Measurement* – The characteristic time for electric-field driven reorientation in the $N_F$ phase is $\tau = \eta/PE$, the intrinsic response time for an induced 90º rotation starting from the high-torque situation where $P$ is normal to $E$ [15]. Here $\eta$ is the orientational viscosity, which is strongly dependent on $T$ in the $N_F$ phase. The driving voltage used in these measurements produces polarization reversal which takes place in RM734 in a time interval $\tau_R \sim 10\tau = 10\eta/PE$ [15]. Polarization reversal takes this long because under conditions of rapid field reversal, $P$ is generally oriented antiparallel to $E$ immediately after the field switches, at a very low-torque orientation through most of the cell volume. We may therefore use $\eta(c,T) = 0.1[P(c,T)\tau_R(c,T)]E$, to obtain an estimate of $\eta(c,T)$. The results are plotted in ***Fig. 5C***. Since the measurements were made at a fixed field amplitude, the $\tau_R$ data are plotted here as the scaled quantity $\tau_R(T)P(T)/P_{sat}$, which, with $P(T) = P_{sat}$ for $T < T_{sat}$ directly displays the viscosity ($\tau_R P/P_{sat} \propto \eta$), over the entire $N_F$ temperature range.

The measured viscosity of all of the mixtures is $\eta \sim 0.05$ Pa·s at the highest temperatures in the $N_F$ phase, increasing on cooling to $\eta \sim 3$ Pa·s at the longest measurable times (when $\tau_R$ reaches 10 msec). The viscosity of each mixture shows a nearly Arrhenius-type dependence on temperature (***Fig. 5C***), suggestive of a barrier-limited dissipation process. The experimental data do generally exhibit upward curvature, trending above the Arrhenius line at the lowest temperatures, which we attribute to the approach to a transition to a glassy state. An onset temperature, $T_g(c)$, taken to be where the measured polarization has dropped to 80% of $P_{sat}$ ($f(T) = 0.8$), which is coincidentally also where $\tau_R \sim 5$ msec, is shown as open triangles in ***Fig. 2***. This transition temperature varies nearly linearly with concentration, paralleling that of the $T_{NF}$ transition.

*Room Temperature Ternary Mixture* – The low-temperature dynamics of the c = 90 wt% DIO mixture, whose $N_F$ phase persists to room temperature, were explored further by mixing in a third component, W1027 (shown in ***Fig. S2***), to make a (70 wt% DIO)/(15 wt% RM734)/(15 wt% W1027) mixture. Samples of this mixture formed a room temperature, fluid $N_F$ phase that was stable against crystallization for many hours. The temperature dependence of the viscosity of this mixture is plotted as white circles in ***Fig. 5C***.

*Phase behavior* – In the case of DIO/RM734, we observe binary mixtures that exhibit a first-order transition between two phases that span the phase diagram across all DIO concentrations $c$. Such mixtures are considered "ideal" if, in the calculation of the phase boundary temperature $T(c)$, the entropy of mixing is the only specifically mixing-related thermodynamic contribution that needs to be considered, besides a linear weighting of the transition enthalpy ($\Delta H$) and entropy ($\Delta S$) change of the individual components, based on their mole fraction. This is to say that "excess" contributions to the difference of Gibbs potential between the two phases, $\Delta G$ are negligible. Such contributions would appear in an A/B mixture, for example, if there were attraction, repulsion, disordering, or ordering in A-B molecular pairing that differed from the simple averaging of these effects in A-A and B-B pairing. Under conditions of ideal phase behavior, $T_{NF}(c)$, the center temperature of the phase coexistence range at the transition to the $N_F$ phase, is described by the Schroeder – van Laar (SvL) equations [25,26]:

$$T_{NF}(x) = \frac{\Delta H_{ave}(x)}{\Delta S_{ave}(x)} = \frac{x T_{DIO} \Delta S_{DIO} + (1-x) T_{RM734} \Delta S_{RM734}}{x \Delta S_{DIO} + (1-x) \Delta S_{RM734}}$$

where $x$ = mole fraction, $T_{DIO}$ = 343 K and $T_{RM734}$ = 405 K are the neat DIO and RM734 transition temperatures to the $N_F$, and $\Delta S_{DIO}$ = 0.07$R$ [6], $\Delta S_{RM734}$ = 0.06$R$ [9], $\Delta H_{DIO}$ = 0.2 kJ/mol [6], and $\Delta H_{RM734}$ = 0.2 kJ/mol [9] are the per-mole quantities of the pure components. This theoretical SvL phase boundary is generally curved in the $x,T$ plane, but is confined in temperature to the range $T_{DIO} < T(x) < T_{RM734}$. Under ideal mixing conditions $\Delta S(x)$ should linearly interpolate between these limits. Inspection shows that if $\Delta S_{DIO} = \Delta S_{RM734}$, then the condition in the RM734/DIO mixtures is obtained: $\Delta S(x)$ is constant and is eliminated from the equation, with $T(x)$ becoming a linear function of $x$, and the phase boundary forming a straight line between $T_{DIO}$ and $T_{RM734}$ across the phase diagram in $x$, and an almost straight line in $c$ because of the small difference in molecular weights ($MW_{RM734}$ = 423, $MW_{DIO}$ = 510). DSC measurement of $\Delta S(x)$ at the intermediate concentrations gives $\langle \Delta S(x) \rangle$ = 0.056±0.02 $R$, comparable to the values of $\Delta S_{DIO}$ and $\Delta S_{RM734}$ given earlier in this paragraph.

At the RM734 end of the phase diagram, the N – $N_F$ transition is direct and first order, whereas at the DIO end, the phase sequence involves two first-order transitions: N – $SmZ_A$ – $N_F$. However, the linear variation of $T(x)$ vs. $x$ is maintained irrespective of whether the transition into the $N_F$ is from the N or the $SmZ_A$ phase. Since $T(x)$ is governed by the intersection of the Gibbs free energy surfaces governing the N – $N_F$ transition, which linearly interpolate between those of the pure components, the linearity of $T(x)$ suggests that the thermodynamic effect of the N – $SmZ_A$ transition is minor. This is likely a consequence of the extremely small N–$SmZ_A$ transition enthalpy ($\Delta H_{NZ}$ = 0.003 kJ/mol [8]), and is also consistent with the $\langle P = 0 \rangle$ nature of both the paranematic N and antiferroelectric $SmZ_A$ phases: all of the net polarization of the $N_F$ phase is developed through the final transition to the $N_F$, at the phase boundary marked with magenta dots in **Fig. 2**. The small $\Delta H_{NZ}$, the linearity of the dependence of $T_{NF}(c)$ vs. $c$, and the similarity across the phase diagram of the transition entropy of the final transition to the $N_F$ are consistent with the $P(T,c)$ curves having comparable saturation values.

The current proposed models for the phase change from the quadrupolar but nonpolar N phase to the quadrupolar and polar $N_F$ phase are that it is either a first-order Landau-de Gennes mean-field transition [4,5], or an Ising-like orientational transition of molecular dipoles having a binary choice of orientations (along $+n$ or $-n$), made first-order by long-range dipole-dipole interactions [7]. In both cases, the polarization $P(T)$ is the principal order parameter of the transition. **Fig. 5B** shows that the growth of $P(T)$ is similar for the different concentrations, and **Fig. 5D** that $P_{sat}(c)$ has a weak linear dependence on $c$. In the context of ideal mixing, this observation constrains the dependence on $c$ of the parameters in such theories of the transition. In the Ising-like system, for example, since $T_{DIO}$ is somewhat smaller than $T_{RM734}$, the Ising interaction energy $J_{ij}(x)$, which gives the local ferroelectric interaction between pairs of dipoles ($i,j$ = DIO,DIO; $i$ = DIO, $j$ = RM734; $i,j$ = RM734,RM734), and which is proportional to $T(x)$, must linearly interpolate like $T_{NF}(x)$. This happens only if $J_{DR} = (J_{DD} + J_{RR})/2$, for which condition $\Delta G$ will have no "excess" internal energy. For the nearest-neighbor Ising model (giving a second-order phase transition in 3D), the entropy is a universal function of $T/J$, in which case there will also be no "excess" entropy contribution to $\Delta G$.

However, the transition to the $N_F$ has been found to be first-order and mean-field-like [4,5], and exhibit highly anisotropic orientational correlations in the N phase [7], features which can be understood with a model that includes the effects of the long-range dipole-dipole interactions on the fluctuations in the N phase. The critical behavior of Ising systems with long-range interactions has been studied extensively in the context of certain magnetic materials that have short-range ferromagnetic exchange forces, but where the long-range dipolar interactions are also important [27,28,29]. Renormalization group analysis shows that the long-range interactions make the magnetic correlations dipolar-anisotropic near the transition in the high temperature phase [30,31], as observed in RM734 [4,7], extending them along $n$, the $z$ axis, by strongly suppressing longitudinal charge-density fluctuations, $\partial P_z/\partial z$ [27,28]. Specifically, starting with the free energy expression Eq. (1) from [5] and adding a dipole-dipole interaction term [7], the structure factor for Ornstein-Zernicke polarization fluctuations of $P_z$ about $q = 0$ becomes $\langle P_z(q)P_z(q)^*\rangle = k_BT\chi(q)$, with $\chi(q) = 1/[\tau(T)(1 + \xi(T)^2q^2) + (2\pi/\varepsilon)(q_z/q)^2]$. Here the correlation length $\xi(T)^2 = b/\tau(T)$, $\tau(T) \propto (T-T_{NF})/T_{NF}$, $b$ is a constant, and $q = q_z + q_Y$. The dipole-dipole (third) term produces extended correlations that grow as $\xi(\tau)$ along $x$ and $y$ but as $\xi(\tau)^2$ along $z$ [28], suppressing $\chi(q)$ for finite $q_z$ as is observed qualitatively from the image sequences of the textures upon passing through the phase transition, and from their optical Fourier transforms (see **Fig. S9** in Ref. [7]). Because of this anisotropy, the correlation volume in this model grows in 3D as $V \sim \xi(\tau)^4$ rather that the isotropic $V \sim \xi(\tau)^3$, reducing the upper marginal dimensionality of the transition to three and making the transition mean-field-like with logarithmic corrections, rather than fluctuation-dominated with 3D Ising universality [32]. The dipole-dipole term scales as $P^2$, so the nearly equal dipole moments of DIO and RM734, and the nearly equal $J$'s, would tend to make this behavior similar across the phase diagram. Assuming for the moment that $T_{DIO} = T_{RM734}$, and therefore that $J_{DD} = J_{RR}$, the "ideal" mixture averaging condition $J_{DR} = (J_{DD} + J_{RR})/2$ reduces to $J_{DR} = J_{DD} = J_{RR}$: the molecules behave identically with respect to their pair interaction energy stabilizing the $N_F$ phase.

_Viscosity_ – The measured viscosities were fit to the Arrhenius form $\eta(T) = A\exp[E_\eta/k_BT]$, as shown in **Fig. 5C**, in order to determine the effective barrier height, $E_\eta$. As can be seen from the uniformity of the slopes, $E_\eta$ is essentially independent of $c$, with an average value of $E_\eta = 7800K$ across the phase diagram. The coefficient $A$, in contrast, varies substantially with $c$, behavior which can be quantified by measuring the viscosity vs. concentration at a single temperature, for example 80ºC. The plot in **Fig. 5D** shows logarithmic additivity of $\eta(T)$ for the DIO/RM734 mixture: $\ln[\eta(80ºC)] = (x)\ln[\eta_{DIO}(80ºC)] + (1-x)\ln[\eta_{RM734}(80ºC)]$. A basic understanding of this behavior can be gained by using the combined Cohen-Turnbull free volume [33]/Eyring rate theory [34] model proposed by Macedo and Litovitz [35]. This model is based on Maxwell's intuitive picture [36], or its contemporary embodiments [37,38], in which viscosity $\eta = G/\nu$, the ratio of $G$, a typical elastic modulus for local shear deformation, to $\nu$, the average rate per molecule of randomly occurring, local structural deconfinement/relaxation events. The rate $\nu$ is given by $\nu = \nu_T p$ $= \nu_T p_E p_V$, where $\nu_T$ is a trial frequency, and $p$ the probability of success, a product of the probability $p_E = \exp(-E_\eta/E)$ that sufficient energy, $E$, will be available [34], and $p_V = \exp(-V/V_f)$ [33], the probability that sufficient free volume, $V$, will be available, where $V_f$ is the average free volume per particle. These probabilities relate viscosity respectively to temperature and density, giving the generalized relationship, $\eta(T) = G/(\nu_T p_E p_V) = (G/\nu_T)\exp(cV_o/V_f + E_\eta/k_BT)$, where $V_o$ is the close-packed volume per particle, and the available energy is on average $k_BT$. Applying this to the DIO/RM734 mixtures, we can consider $G$ and $\nu_T$ to be the same for the two components and, since the Kelvin range is rather narrow, also to be independent of temperature, with the difference in viscosity of the components being the result of a difference in $V_f$. Generally, in a binary mixture we will have $\nu = \nu_T p = \nu_T(p_{DIO})^x (p_{RM734})^{1-x} = \nu_T[(p_E)_{DIO}{}^x (p_V)_{DIO}{}^x (p_E)_{RM734}{}^{1-x} (p_V)_{RM734}{}^{1-x}]$. This expression can be simplified by noting from the similar slopes of the $\ln\eta(T)$ vs $1/T$ fits in **Fig. 5C** that we can take $E_\eta$ to be the same for the two components, $(E_\eta)_{DIO} = (E_\eta)_{RM734}$, from which simplification we get $(p_E)_{DIO} = (p_E)_{RM734} = p_E = \exp(-E_\eta/k_BT)$. The viscosity of the mixture is then given by $\eta(x,T) = (G/\nu_T)[p_E{}^x p_E{}^{1-x} (p_V)_{DIO}{}^x (p_V)_{DIO}{}^{1-x}]^{-1} = \eta_{DIO}(T)^x \eta_{RM734}(T)^{1-x}$, which predicts logarithmic additivity of the viscosities, behavior that is evident from the experimental data plotted in **Fig. 5D**. In the context of the model above, given that $(E_\eta)_{DIO} = (E_\eta)_{RM734}$, this logarithmic additivity implies that the effective free volume in the mixtures is obtained from a linear combination of $V_o/V_f$ for the two components.

_Enantiotropic $N_F$ and $SmZ_A$ phases_ – All single-component $N_F$ materials reported to date, including RM734 and DIO, are monotropic, with the $N_F$ phase observed only on cooling, implying that the $N_F$ state is thermodynamically metastable relative to the crystalline state. When held at a fixed temperature in the $N_F$ state, such single-component materials eventually crystallize, on time-scales ranging from seconds to days [8]. For practical applications of $N_F$ materials, enantiotropic behavior (i.e., a thermodynamically stable $N_F$ phase) is highly desirable. Mixing of multiple components is a well-established route to achieving enantiotropic behavior in liquid crystals, and, in fact, enantiotropic $N_F$ behavior has been described previously by Mandle and co-workers, in mixtures of homologs of RM734 [9]. In order to study the enantiotropic behavior in mixtures of RM734 and DIO, samples were filled into 8 μm-thick cells in the isotropic phase and cooled to room temperature, where they were left undisturbed for two months. Of all the mixtures exhibiting a glassy state, only the 90% DIO sample showed some recrystallization in parts of the cell after this time. The cells were heated slowly, with no applied field, and the phase behavior observed in the polarized light microscope. The spontaneous polarization was measured as a function of temperature using in-plane fields in a subsequent heating cycle. Based on these observations, we determined that the $N_F$ phase is enantiotropic in RM734/DIO mixtures over a range of compositions from 10% to 80% DIO (**Fig. S8**, **Table S1**). In 10% DIO for example, the concentration that exhibits the broadest enantiotropic $N_F$ temperature range, a thermodynamically stable $N_F$ phase is observed from $T = 97.5$°C to 124.7°C. The SmZ$_A$ phase, which is observed in the heating experiments over a wide range of compositions, is also enantiotropic, in contrast to neat DIO, in which the SmZ$_A$ is monotropic.

## *DISCUSSION*

For the isotropic–nematic liquid crystal transition, commonalities observed in the phase behavior for different molecular species having widely different molecular structure have stimulated and supported the notion that the essential elements of nematic liquid crystal structure and ordering could be modeled based on a few relevant molecular features. Thus, nematics were found to be dielectric and nonpolar in the absence of a field, separated from the isotropic by a first-order phase transition with transition enthalpies ~ 1 kJ/mol, and optically uniaxial, with a birefringence that increased slowly with decreasing temperature or increasing concentration. Maier-Saupe [39] and Onsager [40] showed that anisotropic steric shape and/or van der Waals forces, employed to describe intermolecular interactions in simple mean-field or second virial statistical mechanical models, were the molecular features required to get a basic description of nematic ordering.

The results presented in this paper suggest that a similar distillation might be possible with respect to the ferroelectric nematic phase, showing that the effects of family origin on the interactions of molecularly distinct species leading to the ferroelectric nematic phase can be accounted for by the simplest averaging procedures to get $\Delta H(T,x)$ and $\Delta S(T,x)$ This means, for example, that at low concentration of either of the components, its isolated molecules interact with the sea of the other molecular family in a fashion similar to how they interact with their own kind.

What are the required generic molecular features for nematic ferroelectricity? There are currently ~60 molecules among the two families known to induce nematic ferroelectricity, including recent papers by Li et al. [8], which reports more than forty molecules synthesized from the two families, and Mandle et al., which reports twenty five variants in the RM734 family [11]. *Fig. 1* and *Table S1* of [8] summarize the observed phase behavior of the Li compounds as pure materials, dividing them into three categories: *green (9 molecules)* – exhibiting an enantiotropic $N_F$ phase; *blue (12 molecules)* – exhibiting a monotropic $N_F$ phase that was difficult to study because of rapid crystallization; and *red (21 molecules)* – exhibiting no $N_F$ phase. This impressive exploration of the effects of a variety of substitutions, shows that within these families there is a general tendency to form the $N_F$ phase.

The similar molecular-rod shape and size (~three rings long), and the similarly large molecular dipole moments (~11 Debye) in the families, suggest that these are essential to exhibiting an $N_F$ phase. While this combination may be necessary, it is certainly not sufficient, based on: (*i*) the extensive pre-$N_F$ literature of longitudinally polar LCs [41,] which exhibit only re-entrant nematic/smectic paraelectric or antiferroelectric phases; and (*ii*) the observation that substituting -CN for -NO$_2$ eliminates the $N_F$ phase in otherwise identical molecules, in spite of their comparable dipole moments [8,9]. This latter result may illustrate the importance of details of the observed tendency for electrostatic head-to-tail self-assembly [7].

Mertelj, Mandle, and coworkers have posited [4,5,10,11] that in the RM734 family, a bulky side group like MeO in the ortho-position was required to stabilize the new $N_X$ phase, an antiferroelectric, periodic array of splay stripes, by making the molecules more pear-shaped. More recent papers [7,8,15,24], and the observations reported here, show that the textures of the $N_F$ phase are often not macroscopically modulated or locally splayed on any observable length scale. Furthermore, some of the more recent additions to the molecular pallet that exhibit the $N_F$ phase have side groups in the middle or on the other end of the molecule (e.g., the RM734 family compounds **2a**-**2c** and **3** in [8]), or lack side groups altogether (e.g., many members of the DIO family, and the RM734 family compound **12** in [8], which is the same as compound *10* in [11]).

The nearly ideal mixing behavior of the chemically dissimilar compounds DIO and RM734 is at first blush surprising but suggests that the thermodynamics of mixing in binary mixtures of these materials is dominated by electrostatic interactions and electrostatic intermolecular association. Despite their distinct functional groups and patterns of chemical substitution, DIO and RM734 have similar dipole moments (~ 11 D) and charge distributions characterized by an alternation in the sign of charge along the length of the molecule, features that are strongly correlated with typical pair-association motifs (e.g., head-to-tail 'chaining' and side-by-side 'docking') observed in atomistic simulations of RM734 and related compounds [7,42]. The role of longitudinal charge density modulation in stabilizing the $N_F$ phase has also recently been addressed in theoretical work by Madhusudana [43]. We hypothesize that the near-ideal miscibility of RM734 and DIO derives from their similar molecular shape, charge distribution, and electrostatic interactions, a hypothesis that may be tested by investigating the mixing behavior of analogs of RM734 and DIO having modified intramolecular charge distributions.

$N_F$ materials are highly unusual polar solvents that quite generally induce polar orientational order in dipolar solute molecules, a phenomenon we term 'solvent poling'. The degree of induced polar order can be quite large, as evidenced by the ferroelectric polarization measurements in binary mixtures of RM734 and DIO reported here, which show that an RM734 $N_F$ host imparts nearly perfect order to DIO solute molecules in the limit of low DIO concentration (and similarly for a low concentration of RM734 solute molecules in a DIO $N_F$ host). This solvent poling phenomenon may simply result from orientation of solute electric dipoles in the large (~ $10^9$ V/m) local electric fields present in the $N_F$ host, but, more generally, will depend on details of intramolecular charge distribution and molecular shape. Solvent poling is a facile route to the creation of novel functional materials with optimized materials properties. For example, materials with large second-order nonlinear optical susceptibility may be engineered by solvent poling of high-beta chromophore molecules in $N_F$ hosts.

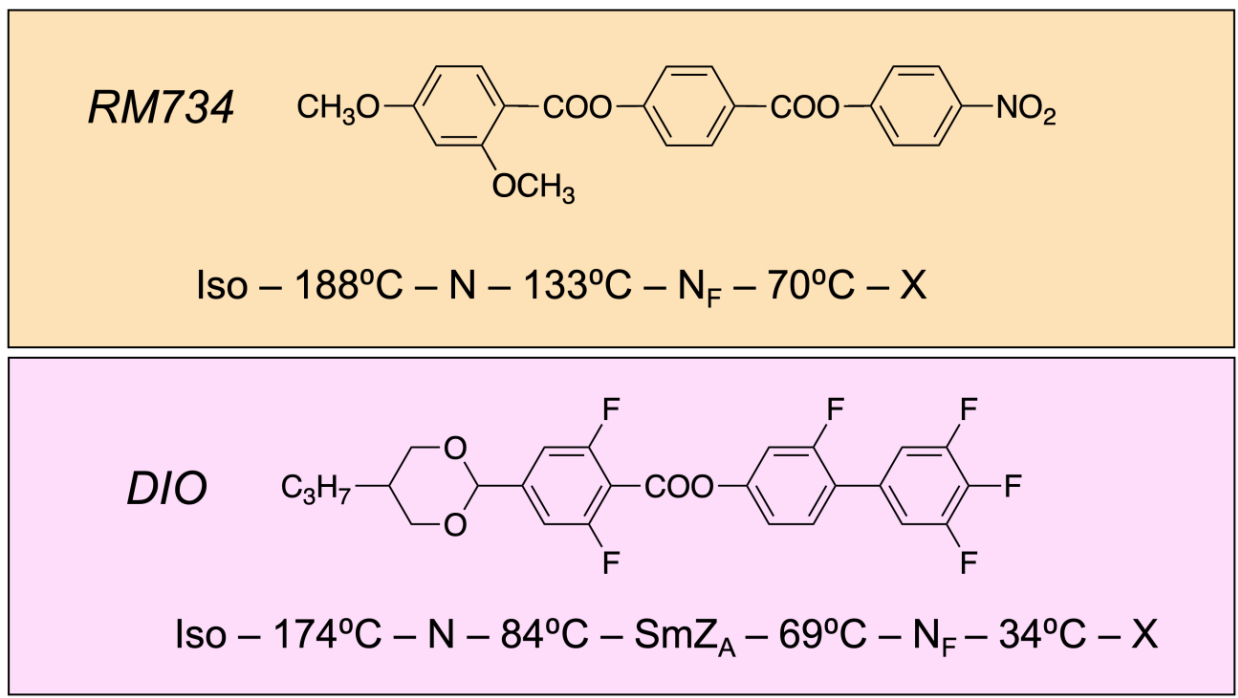

**Figure 1**: RM734 and DIO are representative members of nitro- and fluoro-based molecular families that exhibit novel polar nematic phases. Ferroelectric nematics ($N_F$) have been observed independently in both materials, and in close homologs within each family. DIO exhibits, in addition, an intermediate phase, the $SmZ_A$, recently shown to be an antiferroelectric smectic.

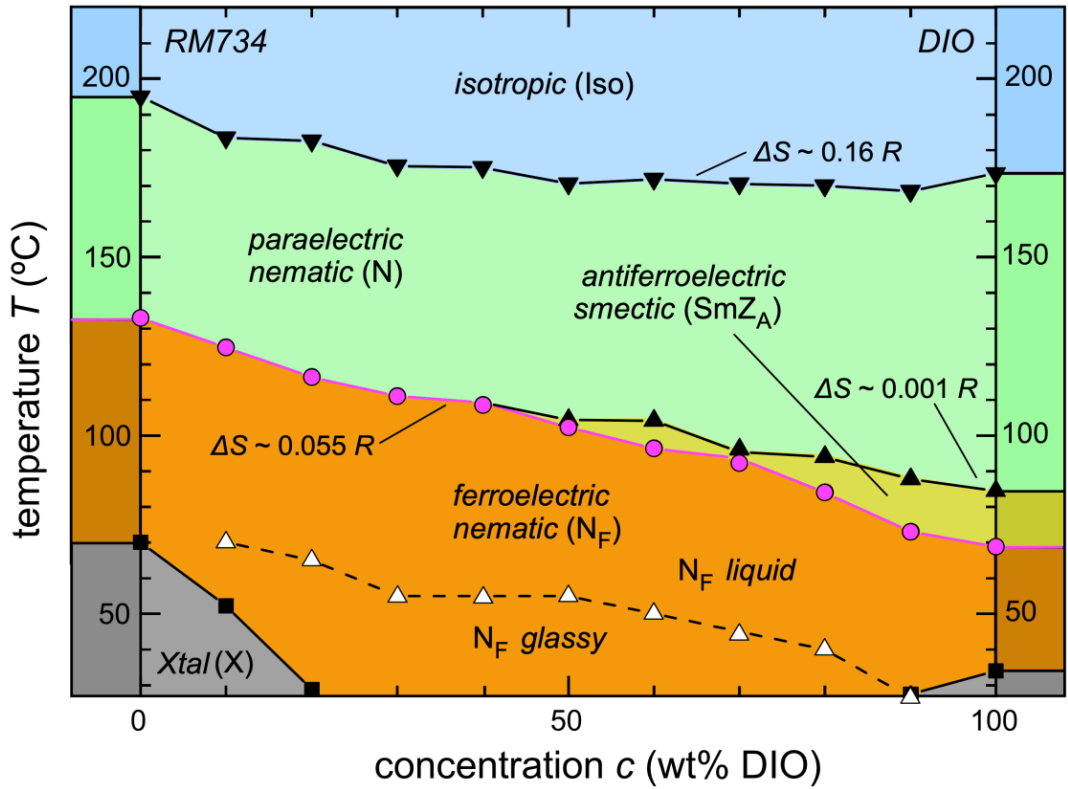

***Figure 2***: Phase diagram of the RM734 and DIO binary mixture. The phase transition temperatures were determined using polarized light microscopy, DSC, and polarization current measurements. Continuous miscibility within the Iso, N, and $N_F$ phases indicates that they are the same in RM734 and DIO. The transitions are first-order, with average entropy changes across the phase diagram of: ▼ − $\Delta S$~$(0.16 \pm 0.01)R$; ● − $\Delta S$~$(0.05 \pm 0.01)R$. The N/SmZ$_A$ transition is ultra-weakly first-order (▲ − $\Delta S$~$0.001R$ for DIO [6]). The temperature of the transition to the $N_F$ phase (magenta circles) varies approximately linearly with concentration, indicating ideal mixing behavior of RM734 and DIO, and equal entropy changes $\Delta S_{DIO} = \Delta S_{RM734}$ at this transition. The SmZ$_A$ phase of DIO, which is not observed for $c < 50$ wt%, has been identified as an antiferroelectric smectic. This phase is lamellar and density-modulated, with the director parallel to the plane of the layers, and the polarization alternating in sign from layer to layer. The effective orientational viscosity of the mixtures, η, increases rapidly on cooling, reaching ~5 Pa · s at the dashed line, heralding the approach to an orientational glass transition. Crystallization is not observed in the glass but occurs at either end of the phase diagram soon (~ 1 hr) after cooling the mixtures into the gray areas.

**Figure 3**: Domain evolution and field response in the N$_F$ phase of neat DIO. This cell is $d = 3.5$ μm thick and has anti-parallel, unidirectional buffing on the two plates, favoring twisted director states. Scale bar is indicated in (*A*). (**A**) The initial director state observed on cooling into the N$_F$ phase is initially uniform-planar (U), like the N and SmZ$_A$ phases, but upon cooling a few degrees further (to $T = 65$ ºC), left- and right-handed π-twist states preferred by the polar surfaces develop. The boundary between domains with opposite twist is a 2π-twist disclination line. Red-to-blue arrow sequences represent the fer-

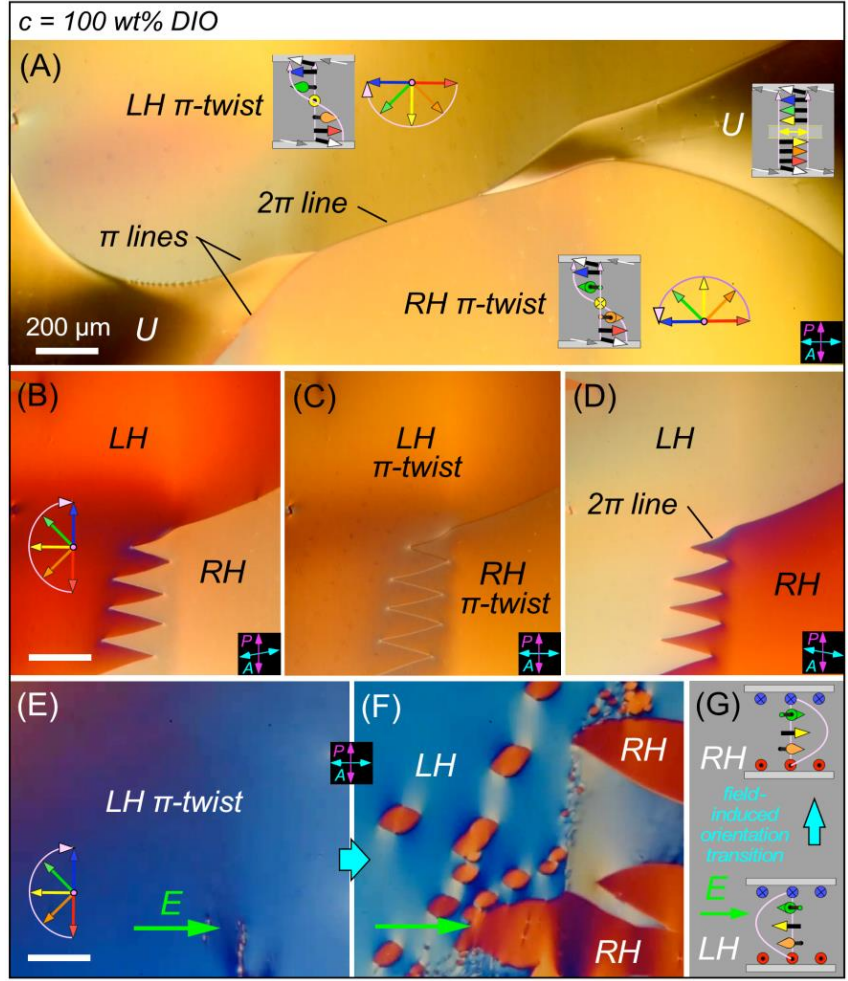

roelectric polarization orientation at increasing heights in the cell ($x = 0$, $d/4$, $d/2$, $3d/4$, $d$), with pink arrows indicating the progression of the polar azimuth $\varphi(x)$ from the bottom to the top of the cell, which is also the light propagation direction. The initial optically uniform (U) state sketched at right, which has a localized polarization reversal wall in the interior of the cell, transitions to a continuous π-twist state by passage of a π-twist surface line. While the uniform state can be extinguished between crossed polarizer and analyzer, the twist state remains birefringent independent of sample orientation. (**B-D**) Decrossing the analyzer lifts the optical degeneracy of the LH and RH states, giving distinct colors that reveal their underlying chirality. Reversing the decrossing angle causes the two twisted states to exchange colors, showing that their director structures are mirror symmetric [15]. (**E-G**) Twist reversal in an applied field. In the relaxed π-twist states, the N$_F$ polarization at mid-height in the cell points in opposite directions in the LH and RH regions. Since the buffing in this cell is parallel to the electrode edges, an applied electric field of particular sign favors either the LH or the RH state and can therefore be used to drive the cell between these two configurations by nucleation and growth of the preferred state. In this example, the field favors the RH twist state. The scale is the same in all images.

*Figure 4*: Textural changes observed on cooling a *c* = 40 wt% DIO mixture through the N – N$_F$ transition in a *d* = 3.5 μm cell with antiparallel, unidirectionally buffed surfaces. (**A**)  In the N phase, the director field is uniform (*U*) and along the buffing direction. (**B**) This condition is initially retained in the N$_F$, once the phase front (dashed white line), characterized by the temporary appearance of irregular, polar domains extended along the director, passes through the cell. (**C**) With a few degrees of additional cooling in the N$_F$ phase, π-twist states stabilized by the increasingly strong antipolar surface anchoring nucleate and grow in the previously uniform N$_F$ region.  Color variations within the twisted regions result from the coupling of the *n(r)*/*P(r)* couple over the area of the domains to in-plane electric fields generated by polarization charge on the twist walls.  Such twisted states are observed in DIO mixtures of all concentrations.  Interestingly, spontaneous periodic modulations of the director like those reported in the N$_F$ phase of neat RM734 [5] are not observed in any of the mixtures or in neat DIO.

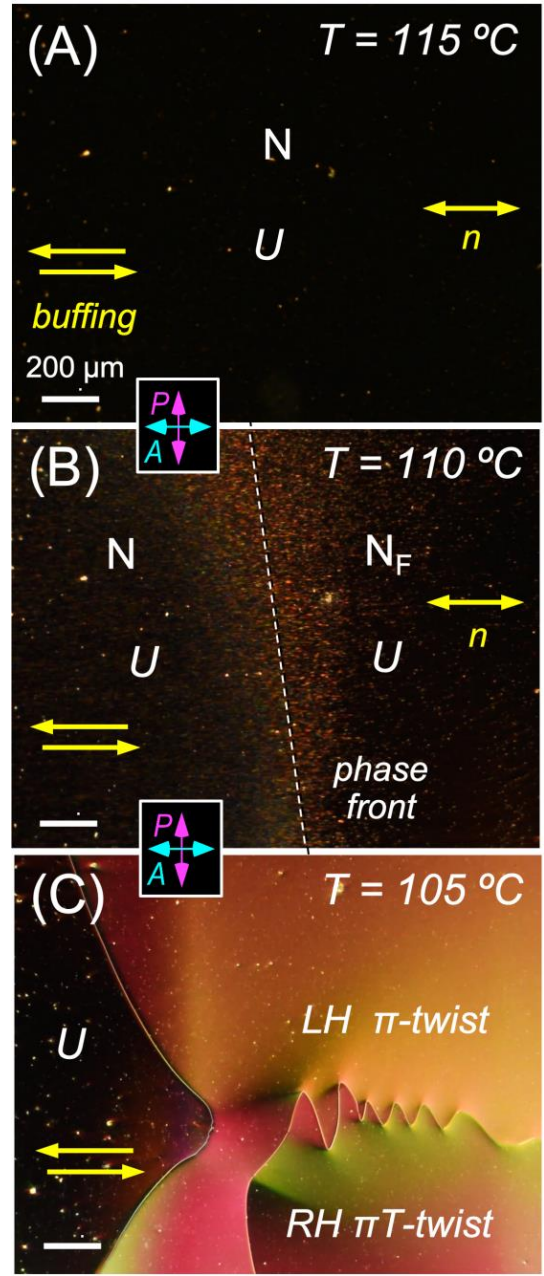

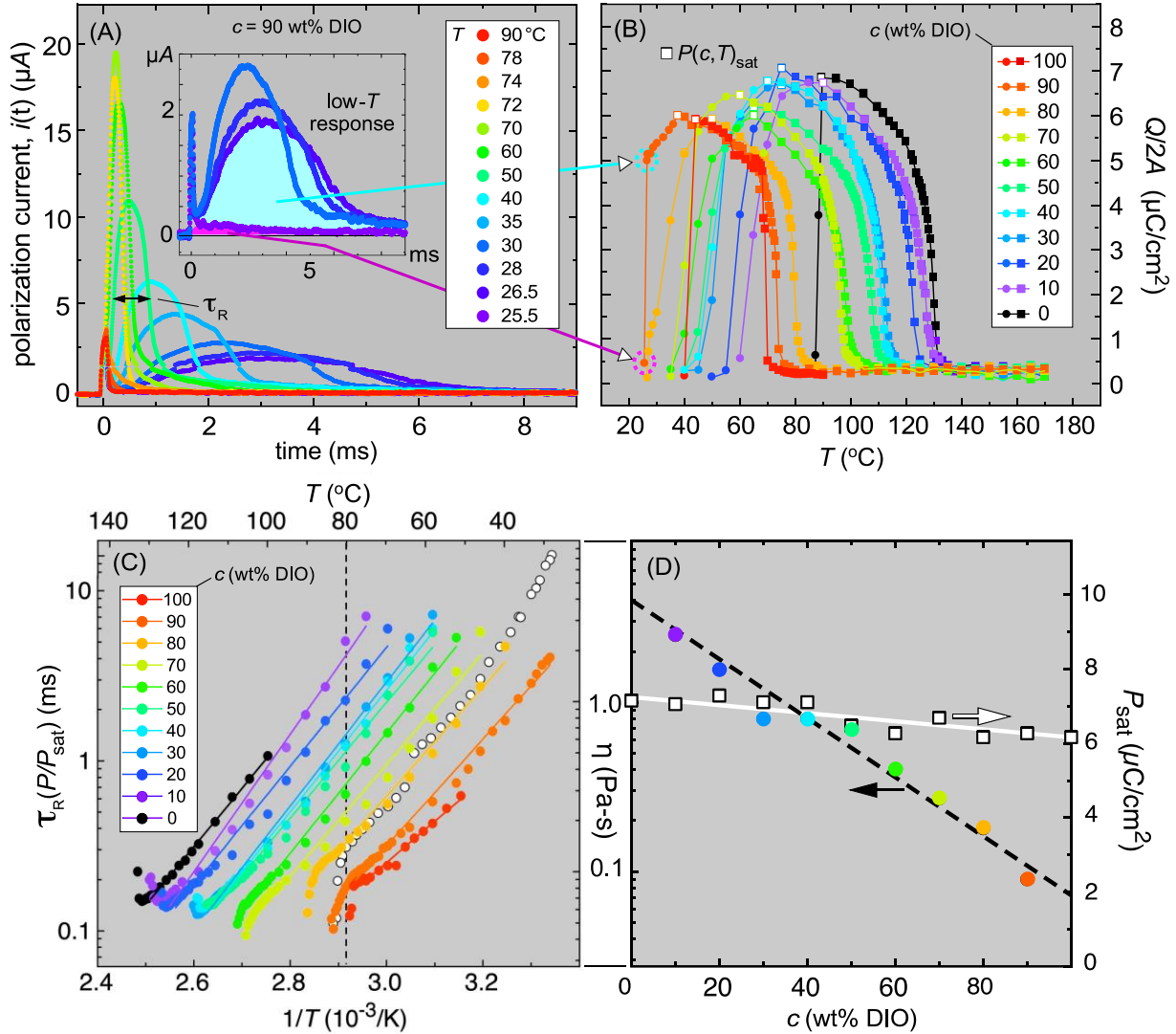

***Figure 5***: Characteristics of polarization reversal in DIO/RM734 mixtures. A 50 Hz square wave with a peak amplitude of 104 V, applied to ITO electrodes with a 1 mm gap in a d = 8 μm thick, planar aligned, cell, results in an in-plane field of 69 V/mm in the center of the gap. This relatively small field is large enough to achieve full reversal of the bulk polarization in the $N_F$ phase at higher temperatures. The polarization is determined by integrating the current through the cell over one half-cycle of the driving voltage (10 msec). (**A**) Polarization current vs time for the c = 90 wt% DIO mixture as a function of temperature. The inset shows the current response at the lowest temperatures. The small current peak seen in the N phase (at $T$=90 ºC), is the capacitive response of the cell to the driving voltage. Upon cooling through the $N_A$ – $N_F$ transition, a large polarization reversal current peak develops, peaking at around $T$ = 70 ºC, before becoming smaller and broader on further cooling due to the increasing viscosity. The polarization reversal time, $\tau_R$, is taken as the full width of the current peak at half-maximum. The inset shows the response at low temperatures. At $T = 26.5$ ºC, the sample still switches ($\tau_R = \sim 6$ msec) but on cooling to $T = 25.5$ ºC, polarization reversal is prevented by crystallization. (**B**) Integrated charge/area, $Q/2A$, transported during field reversal in neat RM734 and DIO and their mixtures. For $T > T_{sat}$ (solid squares), complete switching of $P$ takes place between each field reversal and $Q/2A$ corresponds to the ferroelectric polarization $P(c,T)$. The maximum polarization, $P_{sat}(c)$, indicated with white squares here and in (D), is measured at a temperature $T_{sat}$, below which the director reorientation slows down and the polarization reversal is incomplete (solid circles), so that $Q$ reflects only part of the true polarization density ($Q < 2AP$), which, based on computer simulations and additional measurements, increases weakly from $P_{sat}(c)$ as $T$ is decreased from $T_{sat}$.

(**C**) Plot of $\tau_R(T)$ scaled as $\tau_R(T)P(T)/P_{sat}$, as a function of inverse temperature. This quantity is proportional to the orientational viscosity $\eta(T) = 0.1\ \tau_R PE$, indicated on the right-hand scale. Reversal times were evaluated only at higher temperatures ($T > T_{sat}$), where reorientation was completed between field reversals. The viscosity shows an Arrhenius-like dependence on temperature. The switching times of the three-component DIO/RM734/W1027 mixture described in the text are plotted as white circles. These were measured at low temperature using a longer integration time.

(**D**) Orientational viscosity variation with concentration (colored circles) at fixed temperature [$T = 80$ ºC, dashed black line in (C)]. The viscosities of the two components are seen to add logarithmically in the mixtures. The saturation polarization $P_{sat}(c)$ (white squares) varies approximately linearly across the phase diagram, indicating ideal mixing in the thermodynamics of the transition to the $N_F$ phase.

*Ideal Mixing of Paraelectric and Ferroelectric Nematic Phases
in Liquid Crystals of Distinct Molecular Species*

Xi Chen,[1] Zhecong Zhu,[1] Mitchell J. Magrini,[2] Eva Korblova,[2] Cheol S. Park,[1]
Matthew A. Glaser,[1] Joseph E. Maclennan,[1] David M. Walba,[2] Noel A. Clark[1]

[1]*Department of Physics and Soft Materials Research Center,
University of Colorado, Boulder, CO 80309, USA*

[2]*Department of Chemistry and Soft Materials Research Center,
University of Colorado, Boulder, CO 80309, USA*

*Abstract*

The organic mesogens RM734 and DIO are members of separate molecular families featuring distinct molecular structures. These families, at the present time, are the only ones known to exhibit a ferroelectric nematic liquid crystal (LC) phase. Here we present an experimental investigation of the phase diagram and electro-optics of binary mixtures of RM734 and DIO. We observe paraelectric nematic (N) and ferroelectric nematic ($N_F$) phases in both materials, each of which exhibits complete miscibility across the phase diagram, showing that the paraelectric and ferroelectric are the same phases in RM734 as in DIO. Remarkably, these molecules form ideal mixtures with respect to both the paraelectric-ferroelectric nematic phase behavior and the ferroelectric polarization density of the mixtures, the principal order parameter of the transition. Ideal mixing is also manifested in the orientational viscosity, and the onset of glassy dynamics at low temperature. This behavior is attributable in part to the similarity of their overall molecular shape and net longitudinal dipole moment (~ 11 Debye), and to a common tendency for head-to-tail molecular association. In contrast, the significant difference in molecular structures leads to poor solubility in the crystal phases, enhancing the stability of the ferroelectric nematic phase at low temperature in the mixtures and making possible room temperature electro-optic effects. In the mixtures with excess DIO, an intermediate phase appears via an ultraweak, first-order transition from the N phase, in a narrow temperature range between the paraelectric and ferroelectric nematics.

*Materials & Methods*

<u>Synthesis of DIO</u> – First reported by Nishikawa et al. [1], *DIO* (2,3',4',5'-tetrafluoro-[1,1'-biphenyl]-4-yl 2,6-difluoro-4-(5-propyl-1,3-dioxane-2-yl)benzoate, **Fig. S1,** compound **3**) is a rod-shaped molecule about 20 Å long and 5 Å in diameter, with a longitudinal electric dipole moment of about 11 Debye. The synthesized compound was found to melt at $T = 173.6^{\circ}$C and have an isotropic (I) phase and two additional nematic-like phases. The transition temperatures on cooling were I – $173.6^{\circ}$C – N – $84.5^{\circ}$C – M2 – $68.8^{\circ}$C – $N_F$ – $34^{\circ}$C – X, similar to those reported by Nishikawa.

Our synthetic scheme, shown in **Fig. S1**, is based on a general synthetic reaction. The key intermediate **1** was purchased from Manchester Organics Ltd., UK and intermediate **2** from Sigma-Aldrich Inc., USA. Reactions were performed in oven-dried glassware under an atmosphere of dry argon. Purification by flash chromatography was performed with silica gel (40–63 microns) purchased from Zeochem AG. Analytical thin-layer chromatography (TLC) was performed on silica gel 60 $F_{254}$ TLC plates from Millipore Sigma (Darmstadt). Compounds were visualized using short-wavelength ultra-violet (UV). Nuclear magnetic resonance (NMR) spectra were obtained using a Bruker Avance-III 300 spectrometer. NMR chemical shifts were referenced to deuterochloroform (7.24 ppm for $^1$H, 77.16 ppm for $^{13}$C).

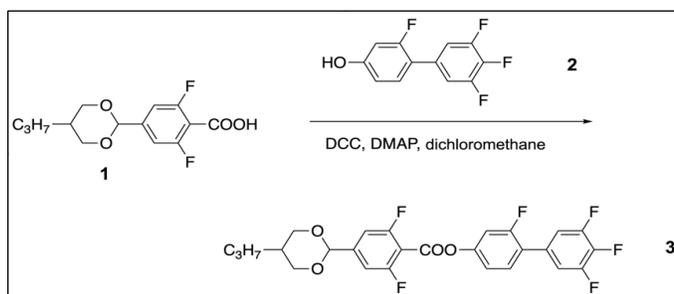

**Figure S1**: Synthesis scheme for DIO.

To a suspension of compound **1** (3.44 g, 12 mmol) and the intermediate **2** (2.91 g, 12 mmol) in $CH_2Cl_2$ (125 mL) was added DCC (4.95 g, 24 mmol) and a trace of DMAP.

The reaction mixture was stirred at room temperature for 4 days, then filtered, washed with water, and with brine, dried over $MgSO_4$, filtered, and concentrated at reduced pressure.

The resulting product was purified by flash chromatography (silica gel, petroleum ether/10% ethyl acetate). The crude product was crystallized by dissolving in boiling 75 mL petroleum ether/20% ethyl acetate solvent mixture, followed by cooling down to -20 °C for 1 hour, yielding 2.98 g (49%) white needles of compound **3**.

$^1$H NMR (300 MHz, Chloroform-*d*) δ 7.64 – 7.35 (m, 1H), 7.24 – 6.87 (m, 6H), 5.40 (s, 1H), 4.42 – 4.14 (m, 2H), 3.54 (ddd, J = 11.6, 10.3, 1.5 Hz, 2H), 2.14 (tddd, J = 11.4, 9.2, 6.9, 4.6 Hz, 1H), 1.48 – 1.23 (m, 2H), 1.23 – 1.01 (m, 2H), 0.94 (t, J = 7.3 Hz, 3H).

$^{13}$C NMR (75 MHz, Chloroform-*d*) δ 162.50, 162.43, 160.85, 159.08, 159.00, 157.52, 150.87, 150.72, 145.60, 145.47, 130.51, 130.46, 117.99, 117.94, 113.22, 113.17, 113.02, 112.93, 112.88, 110.65, 110.30, 110.26, 110.00, 109.95, 98.65, 98.62, 98.59, 72.41, 33.72, 30.05, 19.35, 14.01.

<u>Synthesis of W1027</u> – W1027 (4'-nitro-[1,1'-biphenyl]-4-yl 2,4-dimethoxybenzoate, compound **6**) has a relatively broad nematic phase and is chemically similar to RM734, the difference being that the ester group close to the nitro end of the molecule is replaced by a carbon-carbon linkage, forming a biphenyl structure. The synthesized compound was found to melt at $T = 188°C$ and have isotropic (Iso) and N phases. The transition on cooling were at Iso − 154°C − N −116°C − X.

W1027 was synthesized according to the scheme sketched in *Fig. S2* and described below. Starting materials and reagents were used as purchased from qualified suppliers without additional purification. Intermediates **4** and **5** were purchased from Sigma -Aldrich Inc., USA. 2,4-dimethoxybenzoyl chloride (**4**) (1.41 g, 7.1 mmol) and 4-hydroxy-4'-nitrobiphenyl (**5**) (1.52g, 7.1 mmol) were dissolved in tetrahydrofuran (50 mL), after which triethylamine (0.862 g, 8.5 mmol, 1.2 mL) was added dropwise. The reaction mixture was stirred overnight at room temperature, and extracted with chloroform, washed with water, sodium bicarbonate, and brine, and then dried over MgSO₄. The mixture was then filtered and concentrated at reduced pressure. The crude product was crystallized twice from 150 mL of boiling acetonitrile, affording **W1027** as long pale yellow needles (2.11g. 78%).

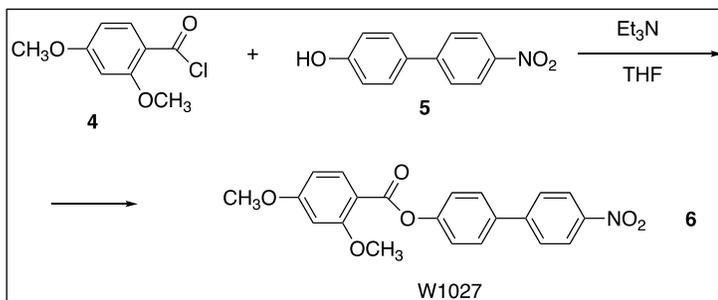

***Figure S2:*** *Synthesis scheme for W1027.*

¹H NMR (300 MHz, Chloroform-*d*)

δ 8.34 − 8.24 (m, 2H), 8.17 − 8.03 (m, 1H), 7.78 − 7.70 (m, 2H), 7.70 − 7.60 (m, 2H), 7.39 − 7.29 (m, 2H), 6.63 − 6.51 (m, 2H), 3.92 (d, *J* = 10.4 Hz, 6H).

¹³C NMR (75 MHz, Chloroform-*d*) δ 165.29, 163.60, 162.47, 151.99, 147.16, 147.03, 136.14, 134.68, 128.50, 127.85, 124.26, 122.91, 110.97, 105.03, 99.16, 56.17, 55.73.

HRMS-EI (m/z) calculated for $C_{21}H_{17}NO_6$ [M-H] ⁺¹, 380.1134; found, 380.1147; devi.: +3.4 ppm.

*Mixtures* – RM734 and DIO were synthesized using respectively the schemes published in [2] and that shown in *Fig. S1*. Samples of the two materials were weighed separately, melted into the isotropic phase, and mixed thoroughly by stirring at 200 °C. The mixtures were studied using standard liquid crystal phase analysis techniques including polarized light microscopy (PLM), differential scanning calorimetry (DSC), and small- and wide-angle x-ray scattering (SAXS and WAXS), as well as combined polarization measurement/electro-optic techniques described previously [3], for establishing the appearance of spontaneous ferroelectric polarization, determining its magnitude, and measuring electro-optic response. DSC studies of RM734, DIO and their mixtures were performed on a Mettler Toledo STARe system calorimeter. The entropy change at the N – SmZ$_A$ transition was too small to be observed.

*Electro-optics* – For electro-optical characterization, the mixtures were generally filled into planar-aligned, in-plane switching test cells with uniform thickness $d$ in the range 3.5 μm < $d$ < 8 μm obtained from Instec, Inc. In-plane indium-tin oxide (ITO) electrodes on one of the glass plates were spaced 1 mm apart. Alignment layers were unidirectionally buffed, antiparallel on the two plates, in a direction nearly parallel to the electrode edges. Such surfaces give quadrupolar alignment of the director along the buffing direction in the N and $N_A$ phases, and polar alignment at each plate in the $N_F$ phase. The antiparallel buffing leads to *ANTIPOLAR* cells in the $N_F$ phase, with a director/polarization field in the plane of the cell but making a $\pi$ twist between the plates [**Error! Bookmark not defined.**]. The splay-bend Freedericksz transition in the N phase was studied using conventional ITO sandwich cells with $d$ = 4.6 μm. Electrooptic measurements were made using a Zeiss polarizing microscope, an EZ Digital FG-8002 function generator, and a Tektronix TDS 2014B oscilloscope. Temperature control was maintained using an Instec HCS402 hot stage.

*Polarization measurement* – The ferroelectric polarization density, $P$, was measured by applying an in-plane electric field to induce polarization reversal while measuring the current flowing through the cell. A 50 Hz, peak voltage $V_P$ = 104 V square-wave voltage was applied across a 1-mm wide electrode gap in the sample cell. The polarization current, $i(t)$, was obtained by monitoring the voltage across a 55 kΩ resistor connected in series with the cell, with results for different temperatures shown in *Fig. 5A*. Polarization reversal was observed to be symmetric, with the +/– and –/+ reversals giving essentially identical current signals. In the isotropic, nematic, and smectic $Z_A$ phases, the current observed following sign-reversal of the applied voltage has a small initial bump and then decays exponentially. This signal corresponds to the RC circuit linear response of the cell and external resistor, giving the initial upward curvature of the measured $P$, due to increasing $\varepsilon$ in the N phase as the $N_F$ is approached in $T$. Upon entering the $N_F$ phase, an additional, much larger current peaks, shown in *Fig. S5*, which come from the field-induced polarization reversal, appear at longer times. The switched net polarization current $Q$ = $\int i(t) dt$ and the corresponding charge density $Q/2A$, where $A$ is the cross-sectional area of the liquid crystal sample in the plane normal to the applied field midway between the two electrodes, are shown in *Figs. 5B, S3*. $Q$ is obtained by integrating this current peak, and the $N_F$ polarization density is given, in general, by $P = Q/2A$, where $A$ is the cross-sectional area of the liquid crystal sample in a plane normal to the field lines midway between the two electrodes.

The data collection system could sample current over a period of 10 msec following a reversal. Since, as *Fig. S5* shows, the current peaks extend to longer times at low temperatures, this finite data collection interval limits the temperature range where $Q = 2AP$. At lower temperatures we have $Q < 2AP$, and decreasing with $T$ as the current peaks get longer (*Fig. 5*). Polarization can also be measured using a triangle wave, which applies less voltage near the +/- reversals, and thus induces a smaller spurious contribution of dielectric polarization near the phase transition (the upward curvature for $T > T_{NF}$). However, with a triangle wave the current peaks are longer than those for the square wave, which in the present case is even more problematical than that just described for the square wave as the viscosity increases with decreasing $T$.

***Visualizing the N–SmZ$_A$ and SmZ$_A$–N$_F$ Phase Transitions***

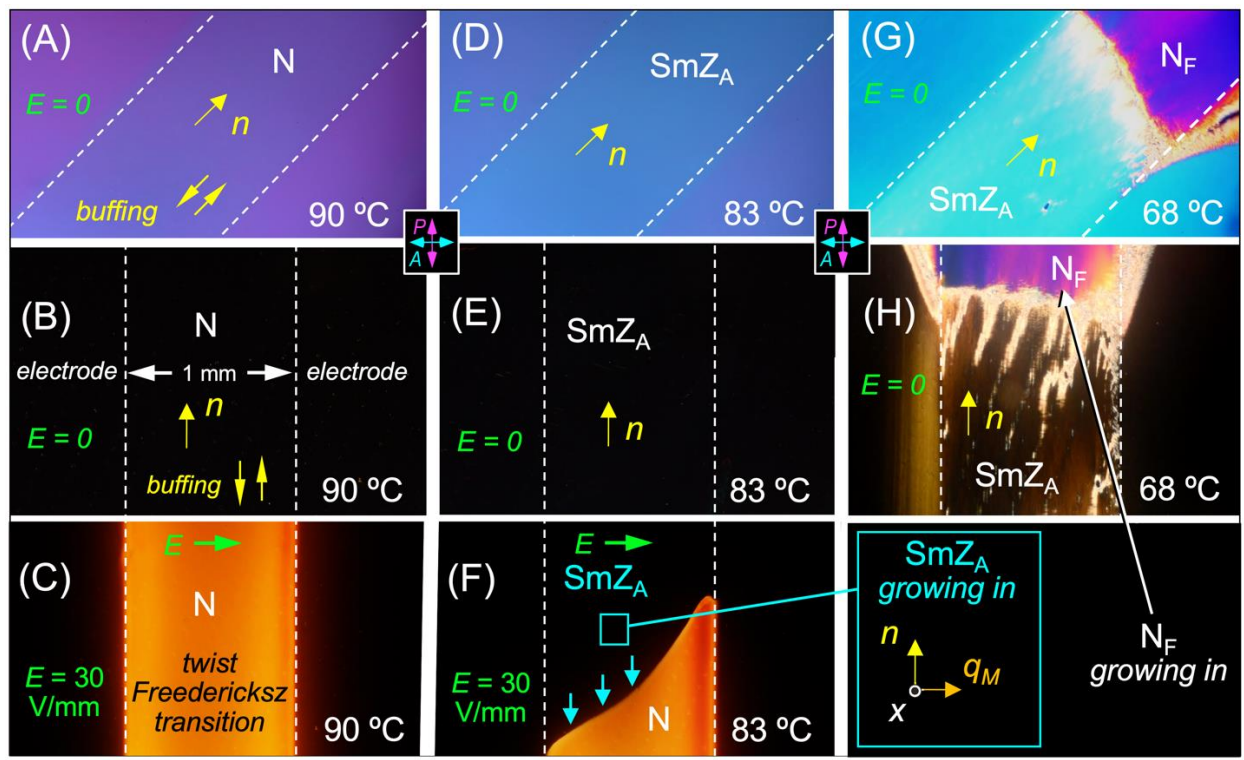

***Figure S3***: Textures of DIO in a $d$ = 3.5 μm cell with in-plane electrodes spaced by 1 mm and with antiparallel buffing parallel to the electrode gap, viewed in PLM with polarizer and analyzer as indicated. (***A,B***) Planar-aligned nematic monodomain with ***n*** uniformly parallel to the plates and the electrode gap, showing (*A*) birefringence when ***n*** is oriented at 45° to the polarizers (the birefringence color corresponds to Δ$n$ = 0.18), (*B*) extinguishing state when ***n*** is oriented parallel to the polarizer. (***C***) Distorted state induced by twist Freedericksz transition when an in-plane *E* field is applied to the previously uniform state. (***D,E***) Planar-aligned SmZ$_A$ monodomain formed on cooling from the nematic, with the smectic layers parallel to ***n*** and normal to the cell plates (bookshelf (BK) geometry). The wavevector $q_M$ is the normal to the smectic layers. The birefringence is slightly larger than in the N phase. Excellent extinction is obtained, as before, when the cell is oriented such that ***n*** is parallel to the polarizer. (***F***) SmZ$_A$ phase growing into the nematic on cooling. An applied electric field induces a distorted twist state in the part of the cell that is in the N phase (as in (*C*)) but in the SmZ$_A$ phase the twist Freedericksz transition is suppressed by the smectic layering. (***G,H***) Twisted N$_F$ state growing into the SmZ$_A$ upon cooling. The antiparallel buffing stabilizes a π-twist state in the polarization-director field which does not extinguish at any orientation between crossed polarizer and analyzer. From [4].

*Freedericksz Transition Threshold Voltages*

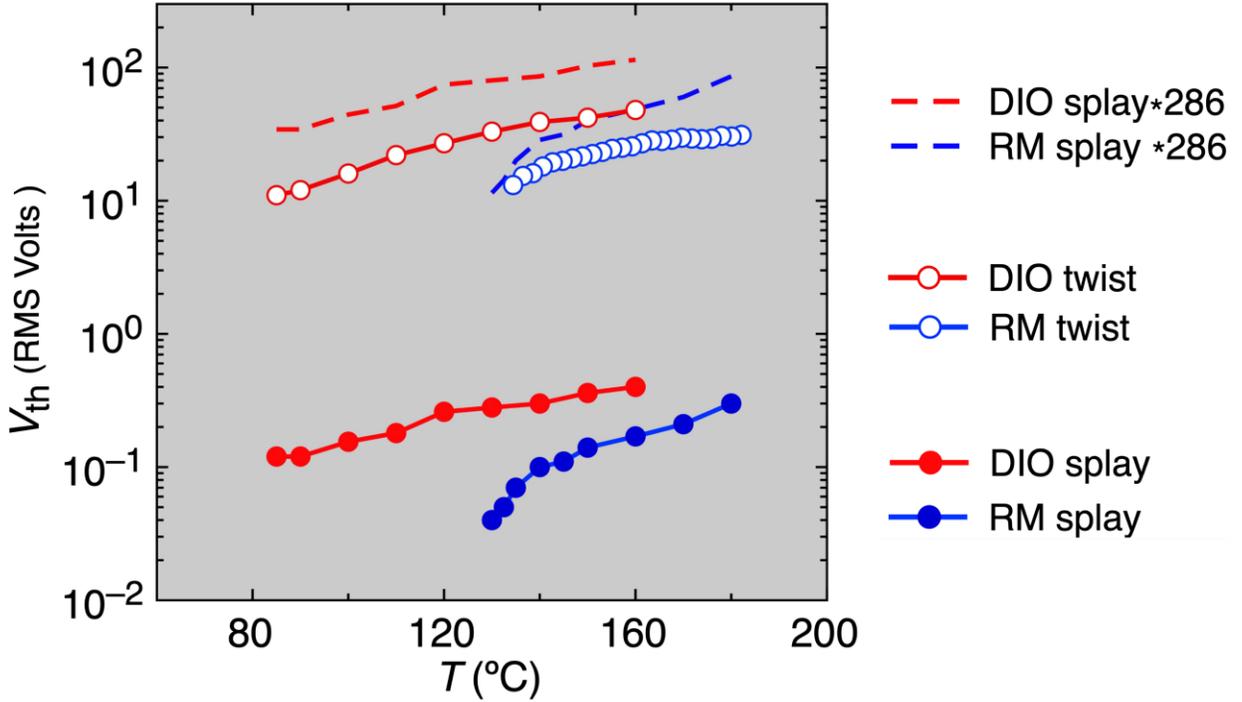

***Figure S4***: Temperature dependence of Freedericksz transition thresholds in the N phase of RM734 and DIO. In-plane applied fields generate twist deformation of the director while fields normal to the cell plates generate splay-bend deformation. The twist transition was observed in 3.5 μm-thick, antiparallel-buffed cells with in-plane electrodes on one surface while the splay-bend transition was observed in conventional ITO sandwich cells with $d = 4.6$ μm. A 200 Hz square-wave field was applied in both cases. The observed splay-bend threshold voltages agree well with $V_{th}^S$ calculated using the $K_S$ and $\Delta\varepsilon$ values given in Refs. [5,6]. Both RM734 and DIO exhibit a general increase in $\Delta\varepsilon$ with decreasing $T$. The Freedericksz thresholds are lower in RM734 and decrease rapidly on approaching the transition to the $N_F$ phase due to the strong pretransitional growth of $\Delta\varepsilon$ [5,6], a phenomenon which does not occur in DIO, where the transition on cooling is to the SmZ$_A$ phase. Comparison of the measurements with the theoretical twist/splay threshold ratio

$$V_{th}^T / V_{th}^S = (D/d)\sqrt{K_T/K_S} = (1\,\text{mm}/3.5\,\mu\text{m})\sqrt{K_T/K_S} = 286\sqrt{K_T/K_S}$$

enables an estimate of the elastic constant ratio $K_T/K_S$. The dashed lines show $V_{th}^S$ scaled up by a factor of 286. This quantity is larger than $V_{th}^T$ in both materials, by a factor of ~3.5 in DIO, indicating $K_S \sim 10\,K_T$ over most of the nematic range.

*Polarization Current Response of the Mixtures*

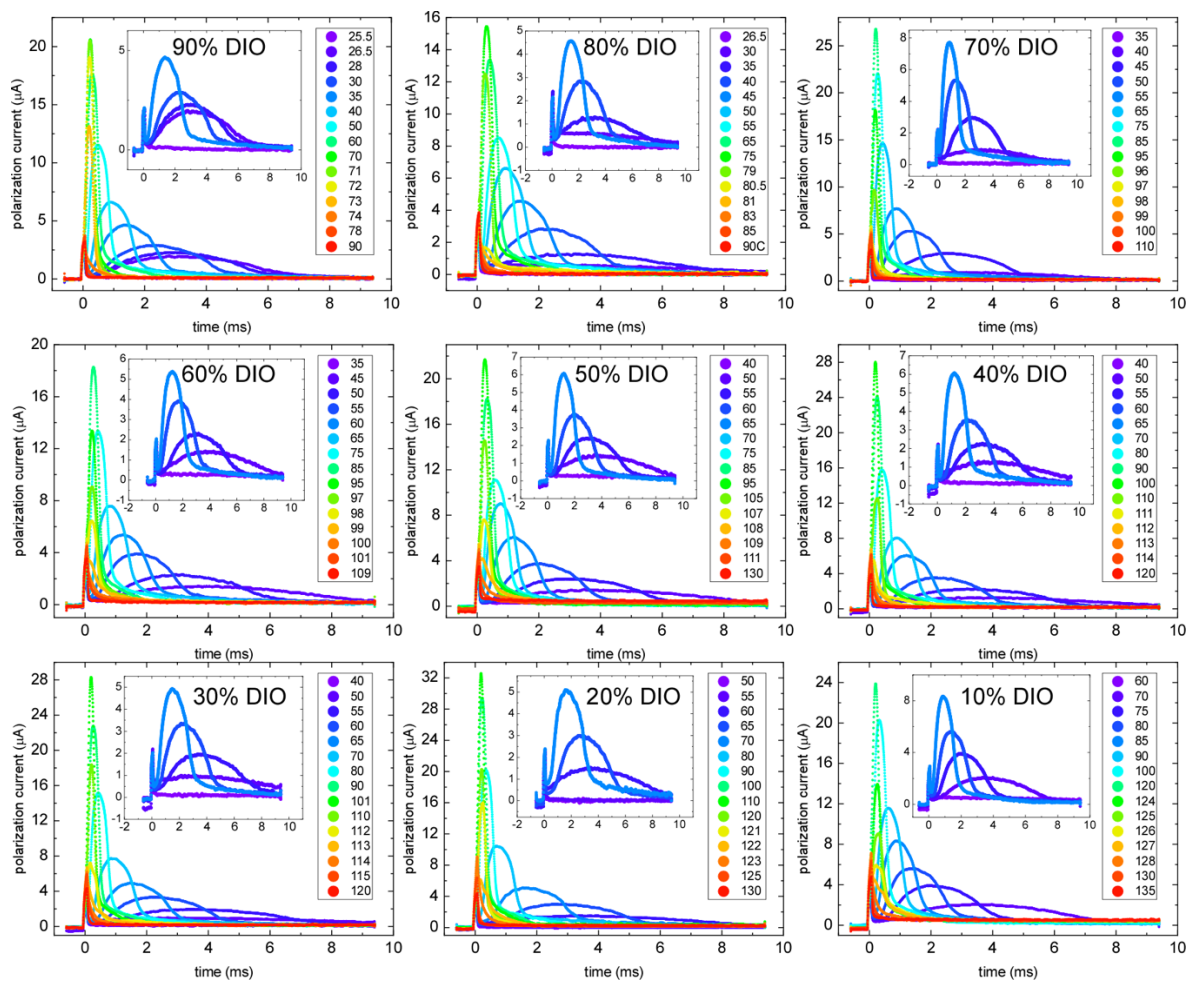

***Figure S5***: Polarization current response to applied field reversal vs. temperature in mixtures with differing DIO concentration. The inset plots show the response at the lowest temperatures. The net charge flow, $Q = \int i(t)dt$, is obtained by integrating the current over the 10 msec half-period of the 50 Hz square wave driving waveform. The corresponding charge density, $Q/2A$, is plotted in ***Fig. 5B***, where $A$ is the cross-sectional area of the liquid crystal sample in the plane normal to the applied field midway between the two electrodes. At higher temperatures in each mixture, where polarization reversal is completed before the next applied field reversal occurs, the polarization, given by $P(T) = Q(T)/2A$, increases on cooling, from a small background value at the N–N$_F$ transition to a value $P_{sat}$ reached at a temperature $T_{sat}$, as seen in ***Fig. 5B***. At temperatures below $T_{sat}$, reorientation of the polarization cannot be completed in the available time and $Q(T)/2A$ decreases, as seen in ***Fig. 5B***. The applied voltage amplitude was 104 V, the electrode gap 1 mm, and the cell thickness 8 µm.

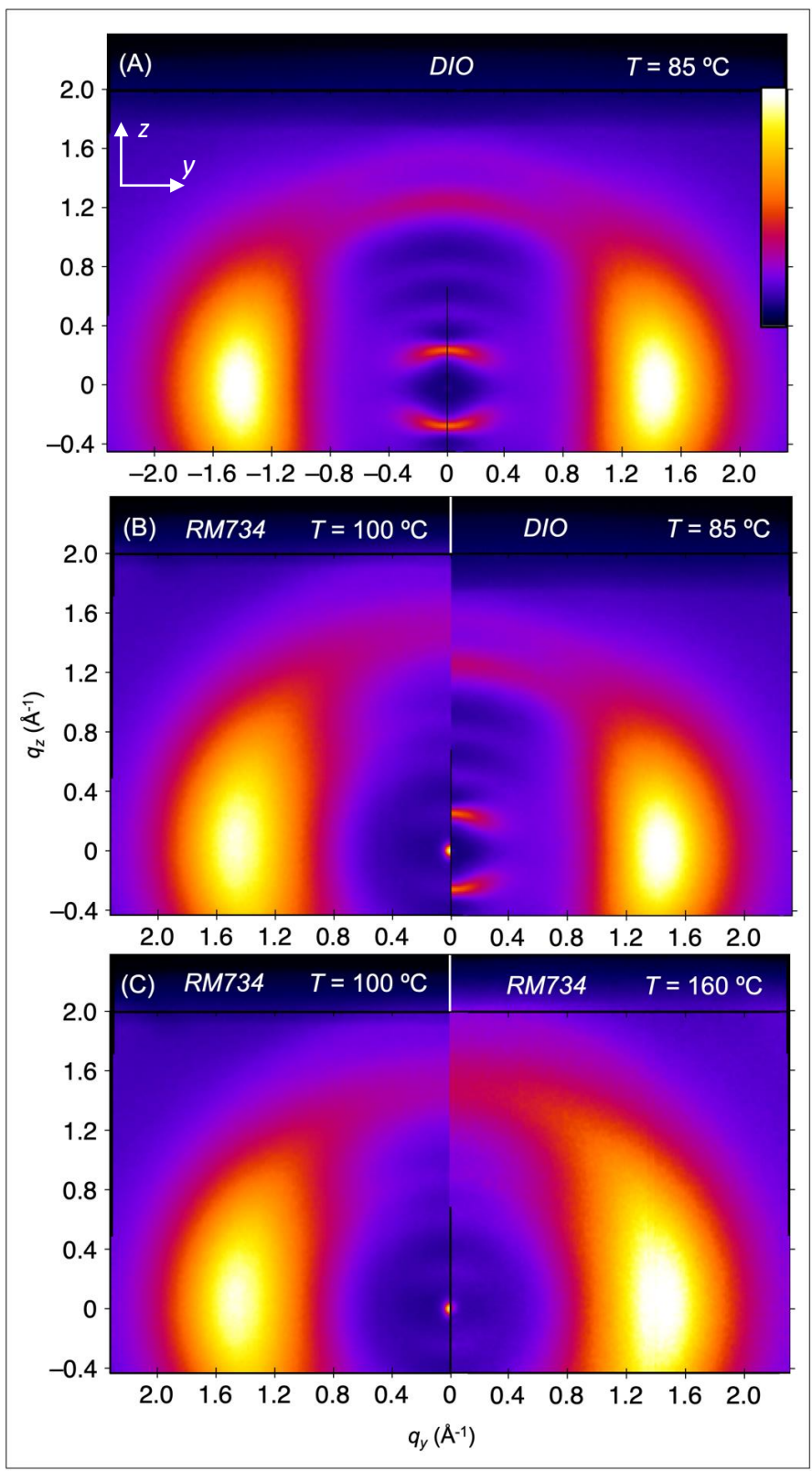

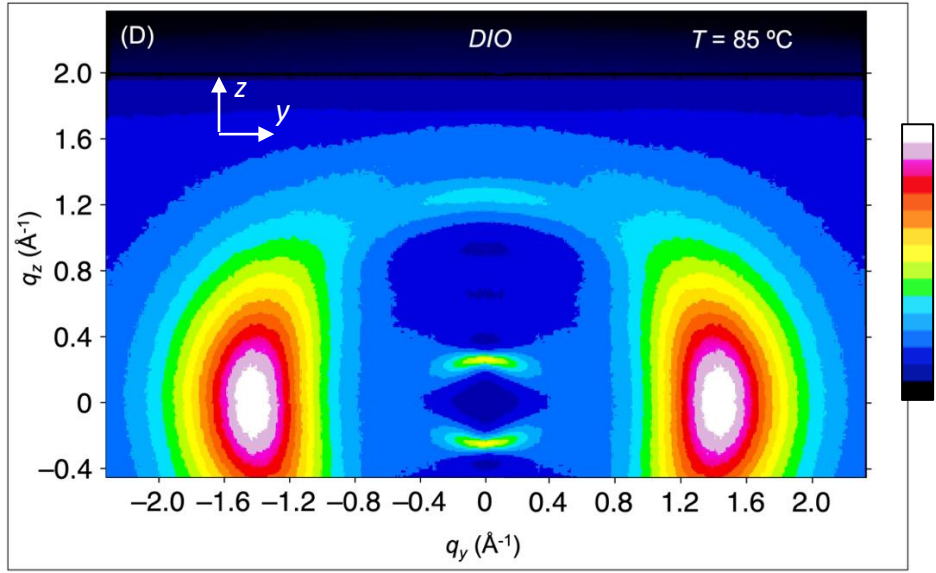

***Figure S6***: Comparison of wide-angle X-ray scattering from RM734 and DIO. The samples have the nematic director magnetically-aligned along ***z*** by a ~1 Tesla magnetic field. The color gamuts are linear in intensity, with the (black) minima corresponding to zero intensity. Scattered intensity line scans, $I(q_z)$, from such WAXS images are shown in ***Figure S7***. Overall, RM734 and DIO exhibit a strikingly similar characteristic WAXS scattering pattern that appears upon cooling into the N phase and does not change very much on cooling into the lower temperature phases, apart from the appearance of the SmZ$_A$ layering peaks (visible in SAXS images [4]). (***A-D***) The WAXS patterns exhibit familiar nematic diffuse scattering features at $q_z$ ~ 0.25 Å$^{-1}$ and $q_y$ ~ 1.4 Å$^{-1}$, arising respectively from the end-to-end and side-by-side pair-correlations, that are typically generated by the steric rod-shape of the molecules and are located respectively at ($2\pi$/molecular length ~ 0.25 Å$^{-1}$) and ($2\pi$/molecular width ~ 1.4 Å$^{-1}$) [1,7]. In contrast to typical nematics, RM734 also exhibits an atypical series of scattering bands for $q_y$< 0.4 Å$^{-1}$ and $q_z$> 0.25 Å$^{-1}$, initially reported in RM734 and its homologs [2,5,8]. Interestingly, DIO presents a qualitatively very similar scattering pattern (*A,B,D*), but with an even more well-defined peak structure, likely a result of the higher variation of excess electron density along the molecule associated with the fluorines. Also notable is that the $q_z$ ~ 0.25 Å$^{-1}$ feature in RM734 is weak compared to that found in typical nematics such as 5CB and all-aromatic LCs [9,10]. We attribute this weak scattering to the head-to-tail electrostatic adhesion in RM734, which makes the molecular correlations along ***z*** more polar and chain-like, reducing the tail-to-head gaps between molecules. The resulting end-to-end correlations are then like those in main-chain LC polymers where there are no gaps and, as a result, the scattering along $q_z$ is weaker than in monomer nematics [11,12,13]. In DIO, on the other hand, the trifluoro group at the end of the molecule generates large electron density peaks that periodically define chain-like correlations along ***z***, even if there are no gaps, resulting in strong scattering at $q_z$ ~ 0.25 Å$^{-1}$. The signal around ***q*** = 0 (for $q < 0.1$Å$^{-1}$) in *B,C* is from stray light.

***Figure S7***:  Line scans, $I(q_z)$, of WAXS images of scattering from DIO and RM734 similar to those in ***Figure S6*** along $q_z$ at $q_y$ = 0.004 Å$^{-1}$. (***A***) Comparison of RM734 at $T$ = 160ºC and DIO at $T$ = 85ºC, the temperatures where the peak structures of $I(q_z)$ are the strongest, reveals common features.  These include the previously reported intense diffuse scattering features at $q_z \sim 0.25$ Å$^{-1}$ and $q_y \sim 1.4$ Å$^{-1}$ [1,7], and the multiplicity of diffuse peaks along $q_z$ previously observed in the RM734 family [2,5,8].  The pairs of similarly colored dots show analogous peaks for the two compounds.  The diffuse peaks located at $q_{zP} \approx 0.25$ Å$^{-1}$ (white dots) correspond to short-ranged order with a quasi-periodic spacing of $2\pi/q_{zP} = p \approx 24$ Å in both materials, comparable to the molecular lengths of DIO and RM734, and, in RM734, to the periodicity of the molecular spacing along the director in head-to-tail assemblies seen in simulations [3].  Viewing such

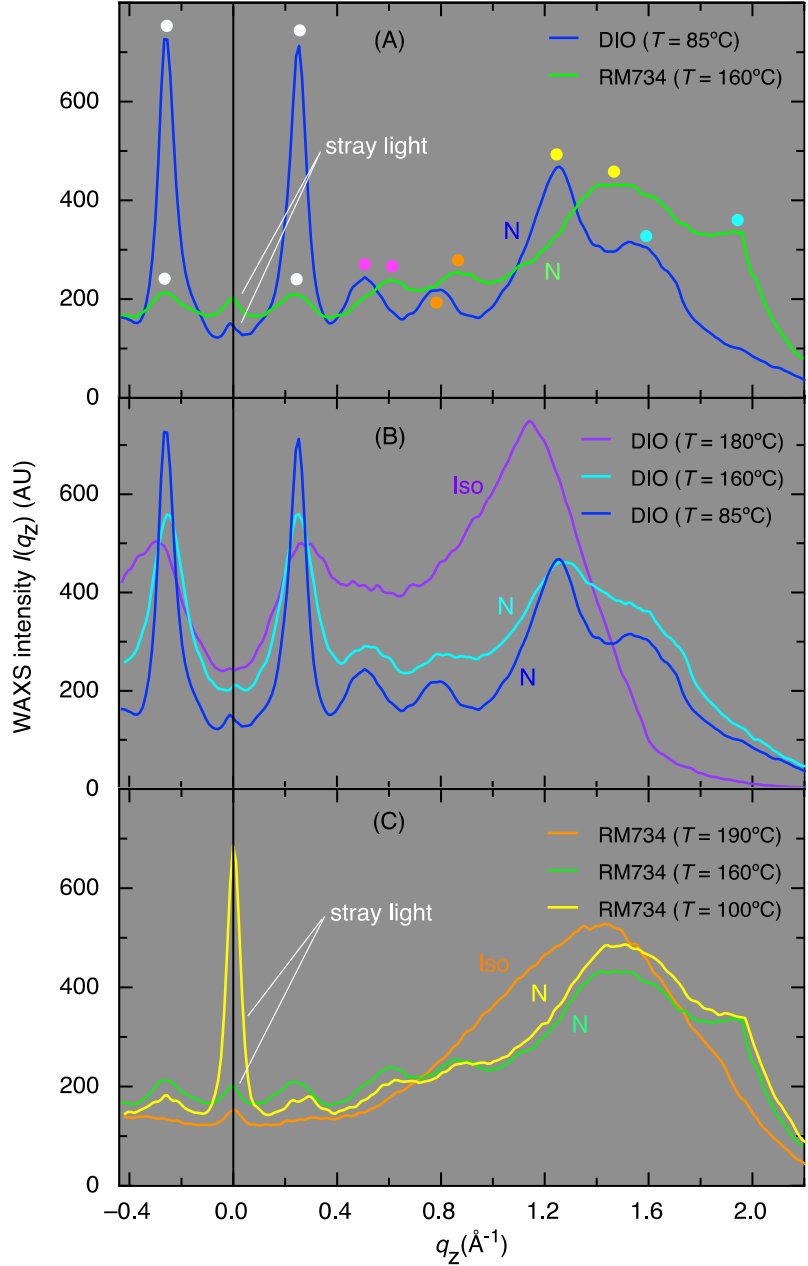

head-to-tail assemblies as one-dimensional chains with displacement fluctuations along the chains, the root mean square relative displacement of neighboring molecules along the chain [see Ref. 14, Supplementary Figure S13], $\sqrt{(\delta u^2)}$, can be estimated from the ratio of the half-width at half maximum of the scattering peak at $q_{zP}$ (0.04Å$^{-1}$) to $q_{zP}$.  This ratio is 0.2, which gives $\sqrt{(\delta u^2)}/p \sim 0.25$ and $\sqrt{(\delta u^2)} \sim 5$ Å.  This is somewhat larger than the rms displacement found in atomistic computer simulations of ~400 RM734 molecules [3], implying that longer length scale fluctuations may also be contributing to the peak width.  (***B,C***) Temperature dependence of WAXS in DIO and RM734. In the N phase of RM734 the peaks in $I(q_z)$ become defined with *increasing* temperature, an unusual behavior in agreement with the results of [7].  In DIO, in contrast, $I(q_z)$ looks the same through most of the N range and below, with the well-defined peaks appearing at $T = 85^{\circ}$C ($B$), and broadening with increasing temperature only near the N–Iso transition, before disappearing in the Iso phase. In the N phase, the sequence of peak positions in **DIO** at $q_z = [0.25, 0.50, 0.78, 1.25, 1.58$ Å$^{-1}$] are in the ratios $q_z/q_{zP} = [1, 2.0, 3.1, 5.0, 7.8]$ which can be indexed approximately as one-dimensional periodicity, giving a harmonic series of multiples of $q_{zP} \approx 0.25$ Å$^{-1}$. In the case of RM734, at $T = 100^{\circ}$C similar indexing of the peak sequence $q_z = [0.28, 0.60, 0.85, 1.46, 1.96$ Å$^{-1}$] is possible, with $q_z/q_{zP} = [1, 2.2, 3.1, 5.3, 7.0]$, but there are significant deviations from harmonic behavior at T $= 160^{\circ}$C, as observed in other members of the RM734 family [6]. In the Iso phase this structure is lost altogether. The nature of the correlations that produce this multiband structure, and their relation to the ferroelectric ordering are currently not understood.

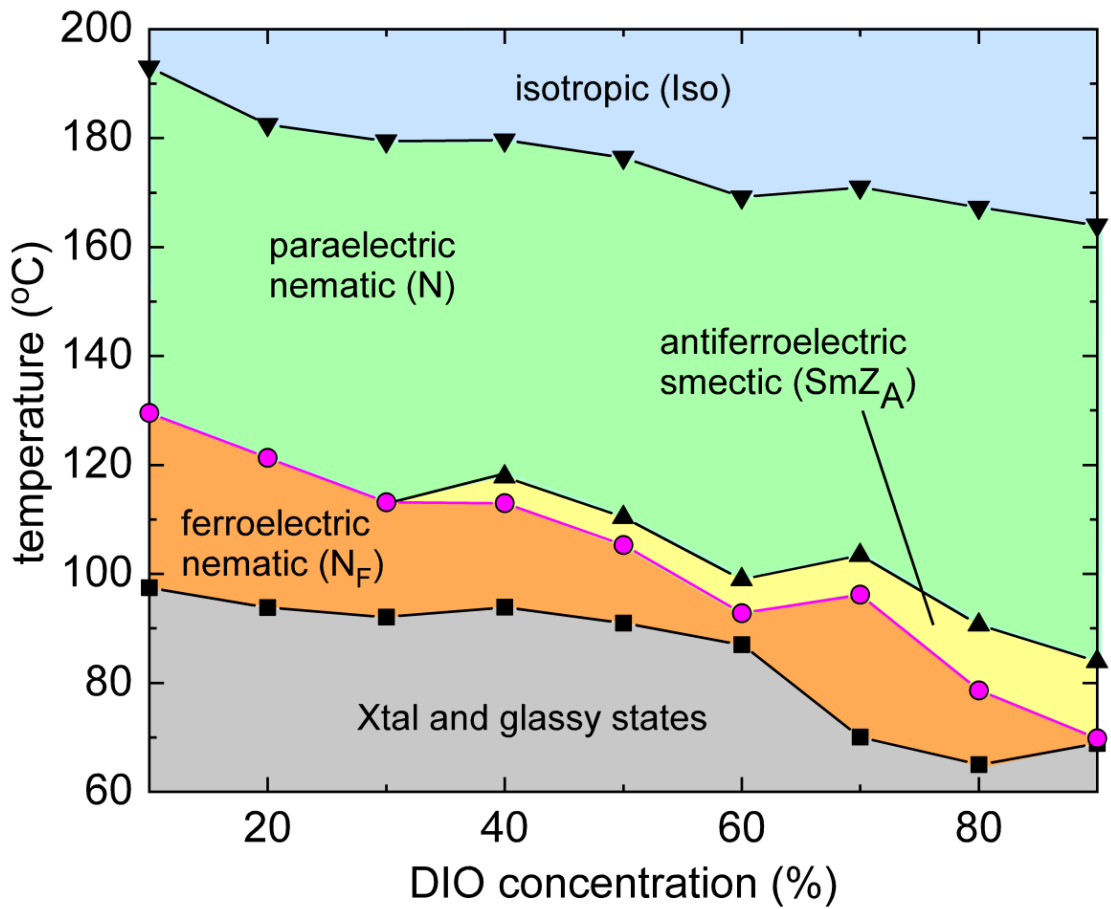

*Figure S8*: Phase diagram of RM734/DIO binary mixtures illustrating their enantiotropic behavior. The transition temperatures were determined on heating using polarized light microscopy and polarization current measurements in cells prepared several months earlier. The $N_F$ phase was observed to be enantiotropic in all of the mixtures. The $SmZ_A$ phase was found to be enantiotropic in mixtures with concentrations in the range of 40 to 90% DIO.

*Table 1*: Upper and lower temperature bounds of the enantiotropic $N_F$ phase depicted in *Fig. S8*.

| DIO% | 10% | 20% | 30% | 40% | 50% | 60% | 70% | 80% | 90% |
|---|---|---|---|---|---|---|---|---|---|
| $T_{upper}$ (°C) | 124.7 | 116.4 | 111 | 109.2 | 102.3 | 92.8 | 93.5 | 78.6 | 69.8 |
| $T_{lower}$ (°C) | 97.5 | 93.8 | 92.1 | 93.9 | 91 | 87.0 | 70.1 | 65 | 68.9 |